\documentclass[10pt]{revtex4}

\usepackage[cp1251]{inputenc}

\usepackage{epsfig}
\usepackage{graphicx}
\usepackage{amssymb,color}

\usepackage{amsmath}
\usepackage{amsthm}
\usepackage{amstext}
\usepackage{comment}
\usepackage{lscape}

\makeatletter \@addtoreset{equation}{section} \makeatother

\begin{document}

\title{ Dirac particle in Newman-Unti-Tamburino spacetime}
\author{\bfseries\itshape
N.G. Krylova$^{1,2}$\thanks{E-mail address: nina-kr@tut.by},
V.M. Red'kov$^{1}$\thanks{E-mail address: v.redkov@ifanbel.bas-net.by}\\
$^{1}$ Institute of Physics, National  Academy of Sciences of Belarus\\
$^{2}$ Belarusian State Agrarian Technical University, Belarus }
\date{}
\maketitle \thispagestyle{empty} \setcounter{page}{1}


\begin{quotation}
We derive the Dirac equation for a particle in the background of the Newman-Unti-Tamburino (NUT) spacetime by applying the tetrad formalism, and separate the angular and radial parts. We get the system of two differential
equations for angular functions and the system of four
differential equations for radial functions. We solve the angular equations in terms of hypergeometric functions and find the NUT-charge dependent  quantization rule for the angular separation constant. As a result of studying the
radial equations, we demonstrate that the probability of particle-antiparticle production on the outer
event horizon decreases with the increase of the NUT charge.
For the massless fermion, we construct the solution of the radial system of Dirac equation in terms of the confluent Heun functions that allows to get the NUT-charge dependent scattering resonances with imaginary energies.
Under the assumption of small NUT charge, we study the
extremal NUT black hole with a single horizon, when  the
Bekenstein-Hawking entropy vanishes identically, and reveal the non-zero NUT charge effects in wave characteristics.

\end{quotation}

\section{Introduction}

The NUT metric has been determined as a vacuum solution of Einstein equations and generalizes the Schwarzschild metric due to the presence of an additional pa\-ra\-me\-ter 
 called the NUT pa\-ra\-me\-ter or the NUT charge
  \cite{Manko2005,Newman,Chakraborty2023}. Currently, the testing of the  Newman-Unti-Tamburino (NUT) charge effects
in the spectra of quasars, supernovae, or active galactic nuclei is one of the striking problems of modern cosmology \cite{Ghasemi-Nodehi2024} as the black holes with NUT
metric are considered to be physically meaningful systems with
some special characteristics. Due to axial symmetry, the black holes with NUT charge may exhibit
some effects demonstrated by the Kerr black holes, such as asymmetry of black
hole shadow or the Lense-Thirring effect  \cite{Chakraborty2023,3}. In \cite{Liu2022},  it has been shown from the thermodynamical analysis that
the observable mass of NUT black hole is  modified by the NUT parameter.
Besides this, the presence of the NUT parameter in the anti-de-Sitter metric leads to the appearance
of a region with the negative Gibbs free energy in  thermodynamics
of black holes associated with  the phase transition
\cite{4}.

To date, the NUT charge is ordinary
referred to as a gravitomagnetic monopole and interpreted as a linear
source of a pure angular momentum (the twisting of the surrounding
spacetime) \cite{Chakraborty2023}.
In  \cite{Atiyah},  it was
shown that the problem of interacting two
Bo\-go\-mol'nyi-Prasad-Sommerfield monopoles may be geometrized, and reduced to finding geodesics in
configuration space 
of the Taub-NUT type.    By this reason, the NUT metric has been actively studied as a monopole-like solution  within the grand unified theories. In this way,  the Kaluza-Klein monopole has been considered as an embedded Taub-NUT gravitational instanton into five-dimensional
theory (the so-called Eucli\-dean Taub-NUT manifold) \cite{4a}. In such a
5-dimensional (5D) model, the Dirac equation has the specific
Kaluza-Klein term
which couples the spin with a magnetic field like in the
Schr\"{o}dinger-Pauli nonrelativistic theory. The
SO(4,1) gauge-invariant theory of the Dirac-field in 5D Taub-NUT
geometry leads to an analytically solvable model which
gives energy levels similar as for
 scalar modes.

The original 4D NUT metric attracted not so much attention of the
scientific commu\-nity. The situation changes in recent years.
In \cite{Godazgar},
  it has been shown that the Taub-NUT metrics may be obtained
from the general class of
asymptotically flat metrics by choosing a gauge field as
corresponding to  the Dirac magnetic monopole. In other words, the
Dirac monopole actually generates  a family of Taub-NUT
solutions. In
general,  other members of this family would cor\-respond to other
stationary  Weyl solutions with the axial symmetry and a non-trivial NUT
charge.

Geodesics in NUT spacetimes were studied  extensively
\cite{Zimmerman1989,Nouri-Zonoz1997,Vandeev2022}.
Howe\-ver, quantum-mechanical problems in the background of NUT spacetime were  investigated insufficiently.
In \cite{Nouri-Zonoz2004}, the Maxwell equations in Taub-NUT space were solved within the Newman-Penrose formalism.
The Dirac equation  in Taub-NUT curved space was considered in the paper \cite{5a}. However, as will be shown below, the variable separation in \cite{5a}  has been performed in a wrong way, and the obtained radial equation system does not have a correct solution of the Dirac equation.

In this paper, we  examine the quantum-mechanical problem of a spin
1/2 particle in the background of the original NUT metric, construct
 analytical solutions of angular and radial systems derived
after separating the variables in Dirac equation and investigate the effects of the NUT charge on the scattering process.

It should be noted that many different approaches exist  for the Dirac equation, even in Minkowski space.
In this connection see, for instance, the detailed reviews are given in \cite{Simulik2025,Simulik2020}.
When working with the Dirac equation in NUT-space, we use the conventional tetrad method developed by Tetrode-Weyl-Fock-Ivanenko in \cite{1928-Tetrode,1929-Weyl,1929-Fock-Ivanenko,1929-Fock(3)}.
The Newman-Penrose
approach \cite{NewmanPenrouse1962} is  the most popular  method to apply the tetrad formalism in the present time.  The difference between these two methods consists  in applying the different
notations. The conventional tetrad approach and the Newman-Penrose formalism are equivalent to each other, we detail this issue  in Supplement A.

\section{ The NUT metric model}

The family of NUT metrics  is defined by the line
element
\begin{equation}
ds^2=\Phi \left( dt - W
d\phi\right)^2-\frac{dr^2}{\Phi}-\left(a^2+r^2\right)
\left(d\theta^2 +  \sin ^2 \theta  d\phi^2\right), \label{nutmetr}
\end{equation}
$$ \Phi = 1-\frac{r_g r + 2 a^2}{r^2+a^2}=\frac{\Delta}{\rho^2}, \, \, W = 2 a (\cos\theta+ C).$$
Here $a$ designates NUT parameter, $r_g$ is a Schwarzschild radius.

The value of the constant $C$ distinguishes between two  main  cases:

\vspace{1mm}
\hspace{-0.4cm} 1) the original NUT-metric at
\begin{equation}
C=-1, \qquad \qquad  W = - 4 a \sin^2 \frac{\theta }{2};
\label{origNUT}
\end{equation}

\hspace{-0.4cm} 2) the Taub-NUT metric at
\begin{equation}
C=0, \qquad \qquad  W = 2 a  \cos\theta.
\end{equation}
\vspace{1mm}

These two cases  describe  different geometries:
the original NUT-metric has the only one singularity at $\theta=\pi$, and the Taub-NUT metric  involves   singularities  at semi-infinite axes
$\theta=0$ and $\theta=\pi.$ From the physical point of view, this two cases should correspond to different physical sources of  NUT-metrics.

Further we employ the original NUT-metric (\ref{origNUT}); the metric tensor reads
$$
g_{\alpha \beta}=\left|
\begin{array}{cccc}
 \Phi & 0 & 0 &  2 a \Phi \left(1-\cos \theta\right)  \\
 0 & -\frac{1}{\Phi} & 0 & 0 \\
 0 & 0 & -a^2-r^2 & 0 \\
 2 a \Phi \left(1-\cos \theta\right) & 0 & 0 & 4 a^2 \Phi \left(1-\cos \theta\right)^2 - \left(a^2+r^2\right) \sin ^2\theta  \\
\end{array}
\right|.
$$

We chose the following tetrad
\begin{equation}
e_{(a)\alpha} (x)=\left |
\begin{array}{cccc}
 \sqrt{\Phi} & 0 & 0 & 2 a \sqrt{\Phi} \left(1-\cos \theta\right) \\
 0 & \frac{1}{\sqrt{\Phi}} & 0 & 0 \\
 0 & 0 & \sqrt{a^2+r^2} & 0 \\
 0 & 0 & 0 & \sqrt{a^2+r^2} \sin \theta \\
\end{array}
\right |.
\label{tetrad}
\end{equation}
Applying the known formulas  \cite{Landau}
\begin{equation}
\begin{split}
\gamma_{abc}=\frac{1}{2}(\lambda_{abc}+\lambda_{bca}-\lambda_{cab}), \quad \\
\lambda_{abc}=(\frac{\partial e_{(a)\alpha}}{\partial x^\beta} -
\frac{\partial e_{(a)\beta}}{\partial x^\alpha} ) e^\alpha_{(b)}
e^\beta_{(c)},
\end{split}
\label{Ricci}
\end{equation}
we calculate the Ricci rotation coefficients:
$$
\gamma_{ab0}=\left |
\begin{array}{cccc}
 0 & \frac{\Phi'}{2 \sqrt{\Phi}} & 0 & 0 \\
 -\frac{\Phi'}{2 \sqrt{\Phi}} & 0 & 0 & 0 \\
 0 & 0 & 0 & \frac{a \sqrt{\Phi}}{a^2+r^2} \\
 0 & 0 & -\frac{a \sqrt{\Phi}}{a^2+r^2} & 0 \\
\end{array}
\right |,  \qquad \gamma_{ab1}=\left |
\begin{array}{cccc}
 0 & 0 & 0 & 0 \\
0& 0 & 0 & 0 \\
 0 & 0 & 0 & 0 \\
 0 & 0 & 0 & 0 \\
\end{array}
\right|,
$$
$$
\gamma_{ab2}= \left|
\begin{array}{cccc}
 0 & 0 & 0 & \frac{a \sqrt{\Phi}}{a^2+r^2} \\
 0 & 0 & \frac{r \sqrt{\Phi} }{a^2+r^2} & 0 \\
 0 & -\frac{r \sqrt{\Phi} }{a^2+r^2} & 0 & 0 \\
 -\frac{a \sqrt{\Phi}}{a^2+r^2} & 0 & 0 & 0 \\
\end{array}
\right|,
$$
$$
 \gamma_{ab3}= \left|
\begin{array}{cccc}
 0 & 0 & -\frac{a \sqrt{\Phi}}{a^2+r^2} & 0 \\
 0 & 0 & 0 & \frac{r \sqrt{\Phi} }{a^2+r^2} \\
 \frac{a \sqrt{\Phi}}{a^2+r^2} & 0 & 0 & \frac{1}{\tan \theta \sqrt{a^2+r^2}} \\
 0 & -\frac{r \sqrt{\Phi} }{a^2+r^2} & -\frac{ 1}{\tan \theta \sqrt{a^2+r^2}} & 0 \\
\end{array}
\right| .
$$

\section{  Dirac equation, separating the variables}

Taking into account the obtained tetrad (\ref{tetrad}) and Ricci rotation coefficients (\ref{Ricci}), the covariant  Dirac  equation for the  fermion of mass $M$
\begin{equation}
\left[i \gamma^a \left( e^\alpha_{(a)} \frac{\partial}{\partial
x^\alpha} + \frac{1}{2} \sigma^{m n} \gamma_{m n a} \right) - M
\right]\Psi=0
\end{equation}
reads
\begin{equation}
\begin{split}
\left[ i \Big ( \gamma^0 \frac{\rho}{\sqrt{\Delta} } + \gamma^3
\frac{2a}{\rho} \sqrt{\frac{1-\cos \theta}{1+ \cos \theta}} \Big )
\frac{\partial}{\partial t}   - i \gamma^1 \Big (
\frac{\sqrt{\Delta}}{\rho}\frac{\partial}{\partial r} +\frac{r
\sqrt{\Delta}}{2 \rho^3} + \frac{\Delta'}{4 \rho
\sqrt{\Delta}}\Big)  \right. \\ \left. +i \gamma^0 \gamma^2
\gamma^3 \frac{a \sqrt{\Delta}}{2 \rho^3}  - i \gamma^2
\frac{1}{\rho} \Big( \frac{\partial}{\partial \theta} +\frac{1}{2
\tan \theta} \Big)  - i \gamma^3 \frac{1}{\rho} \frac{1}{\sin
\theta} \frac{\partial}{\partial \phi} - M \right ]\Psi=0 ,
\end{split}
\label{Dir4dro}
\end{equation}
where we  use the  notations
 $$
 \rho^2=r^2+a^2, \quad \Delta=r^2-r_g
r-a^2, \quad \Phi=\frac{\Delta}{\rho^2},
$$
 and employ the following basis of $\gamma$ matrixes
$$
\gamma^0=\left|
\begin{array}{cc}
0 & I  \\
I & 0 \\
\end{array}
\right| 
, \quad \gamma^1=\left|
\begin{array}{cc}
0 & -\sigma_3  \\
\sigma_3 & 0 \\
\end{array}
\right|
,\quad \gamma^2=\left|
\begin{array}{cc}
0 & -\sigma_1  \\
\sigma_1 & 0 \\
\end{array}
\right|
, \quad \gamma^3=\left|
\begin{array}{cc}
0 & -\sigma_2  \\
\sigma_2 & 0 \\
\end{array}
\right|
, \,
$$
$$
i \gamma^0 \gamma^2 \gamma^3=\left|
\begin{array}{cc}
0 & \sigma_3  \\
\sigma_3 & 0 \\
\end{array}
\right|
, \quad  \Psi=\left|\begin{array}{c}
\xi \\
\chi \\
\end{array} \right|;
$$
$\sigma_i$ designate the Pauli matrices; the bispinor wave
function $\Psi$ consists of two spinor components $\xi, \chi$.

From   eq. (\ref{Dir4dro}), we derive equations in 2-spinor form
\begin{equation}
\begin{split}
 \sigma _1  \left( \frac{ 1}{ \rho } \chi_{,2} + \frac{ 1}{2 \rho \tan \theta
} \chi \right)+ \sigma _2  \left( \frac{1}{\rho \sin \theta}
\chi_{,3} - \frac{ 2 a }{ \rho }  \tan \frac{\theta }{2} \,
\chi_{,0} \right) \quad
\\ + \sigma _3 \left[\frac{\sqrt{\Delta }}{\rho } \chi_{,1}
+\left(\frac{\Delta '}{4 \rho \sqrt{\Delta } }+\frac{\sqrt{\Delta
} }{2 \rho ^3}\rho _- \right)\chi  \right]  + \frac{ \rho
}{\sqrt{\Delta }}\chi_{, 0}  + i M \xi =0,  \end{split}
\end{equation}
\begin{equation}
\begin{split}
 \sigma _1 \left(\frac{1 }{\rho } \xi_{,2} + \frac{ 1}{2 \rho
\tan \theta  } \xi \right) +\sigma _2 \left(\frac{ 1 }{ \rho \sin
\theta} \xi_{,3} - \frac{2   a }{\rho }\tan \frac{\theta }{2} \,
\xi_{,0} \right) \quad \\ + \sigma _3 \left[ \frac{\sqrt{\Delta } }{\rho
} \xi_{,1} + \left(\frac{ \Delta '}{4 \sqrt{\Delta } \rho
}+\frac{\sqrt{\Delta }  }{2 \rho ^3} \rho_+ \right) \xi\right]  -
\frac{ \rho }{\sqrt{\Delta }} \xi_{,0} - i M \chi =0;
\end{split}
\end{equation}
where  we use  the notations
$
\partial_\alpha =\; ,\alpha, \,
\rho_{+}=r + ia, \,  \rho_{-} =r - ia.
$
As the NUT-metric is independent on time and $\phi$, so we can search two
spinors in the form
\begin{eqnarray}
\xi = \Delta^{-1/4} \rho_{+}^{-1/2} e^{-i\epsilon t} e^{im\phi}
X(r ,\theta), \quad  \chi=\Delta^{-1/4} \rho_{-}^{-1/2}
e^{-i\epsilon t} e^{im\phi} Y(r,\theta);
\label{separ1}
\end{eqnarray}
the quantum number
 $m$ represents the third projection of the total angular momentum, correspondingly it takes the values
  $m=\pm1/2, \, \pm3/2, \pm5/2...$\;.
Further we get
\begin{equation}
\begin{split}
\sigma _1 D_\theta Y + i \sigma _2  H Y+ \sigma _3 D_{r-} Y
-\frac{i
\epsilon \rho ^2   }{\sqrt{\Delta }}Y + i M \rho _- X=0, \\
\sigma _1 D_\theta X + i \sigma _2  H X + \sigma _3 D_{r+} X
+\frac{i \epsilon \rho ^2   }{\sqrt{\Delta }}X - i M \rho _{+}
Y=0,
\end{split}
\label{DiracXY}
\end{equation}
where
\begin{equation}
D_{r\pm} =\sqrt{\Delta } \frac{\partial}{\partial r}\pm\frac{i a
\sqrt{\Delta } }{\rho ^2}, \quad
 D_\theta
=\frac{\partial}{\partial \theta} + \frac{1}{2 \tan \theta },
\quad H= \frac{m}{\sin \theta }+ 2  a  \epsilon  \tan
\frac{\theta }{2}.
\label{notDH}
\end{equation}

Applying the following designations for spinors and the Pauli matrices
$$X= \left|
\begin{array}{c}
X_1 \\
X_2
\end{array}
\right|,\quad Y= \left|
\begin{array}{c}
Y_1 \\
Y_2
\end{array}
\right|,\quad
\sigma_1=\left|
\begin{array}{cc}
0 & 1  \\
1 & 0 \\
\end{array}
\right|, \, \sigma_2=\left|
\begin{array}{cc}
0 & -i  \\
i & 0 \\
\end{array}
\right|, \, \sigma_3=\left|
\begin{array}{cc}
1 & 0  \\
0 & -1 \\
\end{array}
\right|,
$$
equations (\ref{DiracXY}) give the system
\begin{equation}
\begin{split}
(D_\theta + H)Y_2 +  \Big( D_{r-} - \frac{i \epsilon
\rho^2}{\sqrt{\Delta}}\Big) Y_1 + i M \rho_- X_1 = 0,
\\
(D_\theta - H)Y_1 - \Big( D_{r-} + \frac{i \epsilon
\rho^2}{\sqrt{\Delta}}\Big) Y_2 + i M \rho_- X_2  = 0,
\\
(D_\theta + H)X_2 +  \Big( D_{r+} + \frac{i \epsilon
\rho^2}{\sqrt{\Delta}}\Big) X_1 - i M \rho_+ Y_1 = 0,
 \\
(D_\theta - H)X_1 -  \Big( D_{r+} - \frac{i \epsilon
\rho^2}{\sqrt{\Delta}}\Big) X_2 - i M \rho_+ Y_2  = 0.
\end{split}
\label{eqX12Y12}
\end{equation}
We will apply  substitutions with the following structure
$$
X_1 = T_1(\theta) R_1(r) , \quad X_2 = T_2(\theta) R_2(r) , \quad Y_1 = T_3(\theta) R_3(r) , \quad
Y_2 = T_4(\theta) R_4(r) ,
$$
then we get
\begin{equation}
\begin{split}
R_4 (D_\theta + H) T_4
+  T_3\Big( D_{r-} - \frac{i \epsilon\rho^2}{\sqrt{\Delta}}\Big) R_3
 + i M \rho_- T_1 R_1= 0,
\\
R_3(D_\theta - H)T_3
- T_4\Big( D_{r-}+ \frac{i \epsilon\rho^2}{\sqrt{\Delta}}\Big)  R_4
+ i M \rho_- T_2 R_2  = 0,
\\
R_2 (D_\theta + H)T_2
+  T_1 \Big( D_{r+} + \frac{i \epsilon\rho^2}{\sqrt{\Delta}}\Big)  R_1
- i M \rho_+ T_3 R_3= 0,
 \\
R_1 (D_\theta - H)T_1
-  T_2 \Big( D_{r+} -   \frac{i \epsilon\rho^2}{\sqrt{\Delta}}\Big) R_2
- i M \rho_+ T_4 R_4   = 0.
\end{split}
\label{eqR1234}
\end{equation}

Let us apply the shortening notations
$$
(D_\theta + H) = a_{+}, \quad (D_\theta - H) = a_{-}, \quad \Big( D_{r-}- \frac{i \epsilon\rho^2}{\sqrt{\Delta}}\Big) =b_{-} \quad
\Big( D_{r-}+ \frac{i \epsilon\rho^2}{\sqrt{\Delta}}\Big) =b_{+},
$$
then
\begin{equation}
\begin{split}
R_4 \; a_{+} \; T_4
+  T_3\;  b_{-} \; R_3
 + i M \; \rho_{-} \; T_1 R_1= 0,
\\
R_3 \;  a_{-}\; T_3
- T_4 \; b_{+}   \; R_4
+ i M \; \rho_{-}\; T_2 R_2  = 0,
\\
R_2 \; a_{+} \; T_2
+  T_1\;  b_{-}^* \;   R_1
- i M \; \rho_{+}\; T_3 R_3= 0,
 \\
R_1 \; a_{-} \; T_1
-  T_2 \; b_{+}^* \; R_2
- i M \; \rho_{+} \;  T_4 R_4   = 0;
\end{split}
\end{equation}
here symbol "$^*$" \hspace{0.5mm}  designates the complex conjugation.

Firstly, we impose constraints
$$
a_{+} T_4 = c_1 T_1, \qquad a_{-} T_1 = c_2 T_4 \ \
\Longrightarrow \ \ a_{-} a_{+} T_4  = c_1 c_2 T_4, \quad a_{+}
a_{-} T_1  = c_1 c_2 T_1,
$$
$$
a_{-}T_3 = c_3  T_2, \quad a_{+} T_2 =  c_4 T_3 \ \
\Longrightarrow \ \ a_{-} a_{+} T_2   = c_3 c_4  T_2, \quad a_{+}
a_{-} T_3   = c_3 c_4  T_3;
$$
In this way, we obtain second order equations for separate functions
$$(a_{-} a_{+}-c_1 c_2) T_4  = 0, \quad (a_{-} a_{+}-c_3 c_4)  T_2   =  0. $$

Let us impose the restriction
\begin{equation}
c_1 c_2 =c_3 c_4,
\label{crestric}
\end{equation}
then functions $T_2$,$T_4$ may differ only by a multiplier,  $T_2=\mu T_4.$
Similarly, from equations
$$(a_{+} a_{-}-c_1 c_2) T_1  = 0, \quad (a_{+} a_{-}-c_3 c_4)  T_3   =  0 $$
it follows the restriction $c_1 c_2=c_3 c_4$, so we get $T_1=\mu_1 T_3.$
Taking this in mind, let us turn to four first-order constraints
$$a_{+} T_4 = c_1 \mu_1 T_3, \quad a_{+} \mu T_4 =  c_4 T_3, \qquad a_{-} \mu_1 T_3 = c_2 T_4, \quad a_{-}T_3 = c_3  \mu T_4;  $$
whence follow the restrictions
\begin{equation}
\frac{c_4}{\mu} = c_1 \mu_1, \quad \frac{ c_2}{\mu_1 }=  c_3  \mu \quad \Longrightarrow  \quad \mu \mu_1
=\frac{c_4}{c_1}=\frac{c_2}{c_3}, \quad \frac{1}{\mu_1}= \frac{c_1}{c_4} \mu = \frac{c_3}{c_2} \mu.
\label{mumu1}
\end{equation}
In this way, we get the constraints
\begin{equation}
T_3= \frac{c_3}{c_2} \mu T_1, \qquad T_4=\frac{1}{\mu} T_2.
\label{T2T3}
\end{equation}
Therefore, we arrive at more simple  substitution (with some freedom in parameters):
\begin{equation}
\Psi=
\left|
\begin{array}{r}
T_1 R_1\\ T_2 R_2 \\   \frac{c_3}{c_2} \mu T_1 R_3 \\  \frac{1}{\mu} T_2 R_4
\end{array}
\right|. 
\label{PsiL}
\end{equation}

Applying the substitution (\ref{PsiL}) to system (\ref{eqX12Y12}),  we get the system with respect to the variables $R_i(r)$:
\begin{equation}
\begin{split}
 \Big( D_{r-} - \frac{i \epsilon \rho^2}{\sqrt{\Delta}}\Big) \mu R_3 +  i M \rho_- \frac{c_2}{c_3} R_1 + c_4 R_4  = 0,
\\
 \Big( D_{r-} + \frac{i \epsilon \rho^2}{\sqrt{\Delta}}\Big) R_4 - i M \rho_- \mu R_2 -  c_3 \mu R_3  = 0,
\\
 \Big( D_{r+} + \frac{i \epsilon \rho^2}{\sqrt{\Delta}}\Big) \frac{c_2}{c_3} R_1 -  i M \rho_+ \mu R_3 +  c_4  \mu R_2  = 0,
 \\
 \Big( D_{r+} - \frac{i \epsilon \rho^2}{\sqrt{\Delta}}\Big)  \mu R_2 + i M \rho_+ R_4 - c_3 \frac{c_2}{c_3} R_1 = 0.
\end{split}
\label{DirR1234}
\end{equation}
In is convenient to transform the system to the new variables:
\begin{equation}
\tilde{R}_1=\frac{c_2}{c_3} R_1, \quad \tilde{R}_3=\mu R_3, \quad \tilde{R}_2=\mu R_2, \quad \tilde{T}_1=\frac{1}{\mu_1} T_1.
\label{TRtilde}
\end{equation}
Then the substitution (\ref{PsiL}) and  system (\ref{DirR1234}) take the form:
\begin{equation}
\Psi=
\left|
\begin{array}{r}
\frac{c_3}{c_2} \,  T_1 \tilde{R}_1\\ \frac{1}{\mu} \, T_2 \tilde{R}_2 \\   \frac{c_3}{c_2}\,  T_1 \tilde{R}_3 \\  \frac{1}{\mu} \,  T_2 R_4
\end{array}
\right|= \left|
\begin{array}{r}
\frac{1}{\mu \mu_1} \,  T_1 \tilde{R}_1\\ \frac{1}{\mu} \, T_2 \tilde{R}_2 \\   \frac{1}{\mu \mu_1} \,  T_1 \tilde{R}_3 \\  \frac{1}{\mu}\,  T_2 R_4
\end{array}
\right| = \frac{1}{\mu} \left|
\begin{array}{r}
\tilde{T_1} \tilde{R}_1\\  T_2 \tilde{R}_2 \\   \tilde{T_1} \tilde{R}_3 \\  T_2 R_4
\end{array}
\right|; 
\label{PsiLt}
\end{equation}
\begin{equation}
\begin{split}
 \Big( D_{r-} - \frac{i \epsilon \rho^2}{\sqrt{\Delta}}\Big) \tilde{R}_3 +  i M \rho_- \tilde{R}_1 + c_4 R_4  = 0,
\\
 \Big( D_{r-} + \frac{i \epsilon \rho^2}{\sqrt{\Delta}}\Big) R_4 - i M \rho_- \tilde{R}_2 -  c_3 \tilde{R}_3  = 0,
\\
 \Big( D_{r+} + \frac{i \epsilon \rho^2}{\sqrt{\Delta}}\Big) \tilde{R}_1 -  i M \rho_+ \tilde{R}_3 +  c_4  \tilde{R}_2  = 0,
 \\
 \Big( D_{r+} - \frac{i \epsilon \rho^2}{\sqrt{\Delta}}\Big)  \tilde{R}_2 + i M \rho_+ R_4 - c_3 \tilde{R}_1 = 0.
\end{split}
\label{eqDRc34}
\end{equation}
In the following, for shortness, the tilde symbols will be omitted.
Taking into account the expressions for operators $D_{r\pm}$  (\ref{notDH}),  we get
\begin{equation}
\begin{split}
   \Big( \sqrt{\Delta }
\frac{\partial}{\partial r}-\frac{i a \sqrt{\Delta } }{\rho ^2} -
\frac{i \epsilon \rho^2}{\sqrt{\Delta}} \Big) R_3   + i M
(r-ia)R_1 = -c_4 R_4 ,
\\
   \Big( \sqrt{\Delta }
\frac{\partial}{\partial r}+\frac{i a \sqrt{\Delta } }{\rho ^2} -
\frac{i \epsilon \rho^2}{\sqrt{\Delta}}\Big) R_2  + i M (r+ia)R_4
= c_3  R_1 ,
\\
    \Big( \sqrt{\Delta }
\frac{\partial}{\partial r}+\frac{i a \sqrt{\Delta } }{\rho ^2} +
\frac{i \epsilon \rho^2}{\sqrt{\Delta}}\Big) R_1   - i M (r+ia)R_3
= -c_4 R_2,
\\
   \Big( \sqrt{\Delta }
\frac{\partial}{\partial r}-\frac{i a \sqrt{\Delta } }{\rho ^2} +
\frac{i \epsilon \rho^2}{\sqrt{\Delta}}\Big) R_4   - i M (r-ia)R_2
= c_3  R_3;
\end{split}
\label{Direqseparation a}
\end{equation}
recall the angular equations (remembering that here $T_1$ designates $\tilde{T_1}$)
\begin{equation}
\begin{split}
(D_\theta + H) T_2   = c_4 T_1 ,  \quad (D_\theta - H) T_1
=  c_3  T_2 .
\end{split}
\label{Direqseparation}
\end{equation}

Instead of two parameter  $c_4$  and $c_3 $, one can preserve only one:
$$
\Big [ (D_\theta + H) (D_\theta - H) -c_3  c_4 \Big ] T_1 =0, \quad - c_3  c_4  =\Lambda^2 ;
$$
$$
\Big [ (D_\theta - H) (D_\theta + H) -c_3  c_4 \Big ] T_2 =0, \quad - c_3  c_4  =\Lambda^2 .
$$
Indeed, let $T_1 =a   \bar{T}_1, \, T_2 = b  \bar{T}_2, $
then  equations (\ref{Direqseparation}) read
$$
(D_\theta + H) b  \bar{T}_2    = c_4  a  \bar{T}_1 ,
\quad (D_\theta - H) a  \bar{T}_1   =  c_3  b  \bar{T}_2.
$$
With the use of new notations
$$
c_4 {a \over b} =  i \Lambda, \quad c_3  {b \over a} = i \Lambda \quad
 \Longrightarrow \Lambda =  \sqrt{- c_4 c_3  },\quad  a = - \sqrt{c_3 },\quad  b = \sqrt{c_4},
$$
we obtain $( \sqrt{- c_3  c_4}= \Lambda)$
\begin{eqnarray}
(D_\theta + H)   \bar{T}_2    =  \Lambda \, \bar{T}_1 ,
\quad (D_\theta - H)    \bar{T}_1   =  \Lambda \,  \bar{T}_2;
\end{eqnarray}
for definiteness, one can choose $c_3 =\Lambda, \,   c_4=-\Lambda$.

Finally,  equations  (\ref{Direqseparation a}), (\ref{Direqseparation}) take the form:
\begin{equation}
\begin{split}
(D_\theta + H) T_2   = -\Lambda  T_1 ,  \quad (D_\theta - H) T_1
=  \Lambda T_2;
\end{split}
\label{DireqAsep_L}
\end{equation}
\begin{equation}
\begin{split}
   \Big( \sqrt{\Delta }
\frac{\partial}{\partial r}-\frac{i a \sqrt{\Delta } }{\rho ^2} -
\frac{i \epsilon \rho^2}{\sqrt{\Delta}} \Big) R_3   + i M
(r-ia)R_1 = \Lambda R_4 ,
\\
   \Big( \sqrt{\Delta }
\frac{\partial}{\partial r}+\frac{i a \sqrt{\Delta } }{\rho ^2} -
\frac{i \epsilon \rho^2}{\sqrt{\Delta}}\Big) R_2  + i M (r+ia)R_4
= \Lambda R_1 ,
\\
    \Big( \sqrt{\Delta }
\frac{\partial}{\partial r}+\frac{i a \sqrt{\Delta } }{\rho ^2} +
\frac{i \epsilon \rho^2}{\sqrt{\Delta}}\Big) R_1   - i M (r+ia)R_3
= \Lambda R_2,
\\
   \Big( \sqrt{\Delta }
\frac{\partial}{\partial r}-\frac{i a \sqrt{\Delta } }{\rho ^2} +
\frac{i \epsilon \rho^2}{\sqrt{\Delta}}\Big) R_4   - i M (r-ia)R_2
= \Lambda R_3.
\end{split}
\label{DireqRsep L}
\end{equation}

Let us remark that according the restriction (\ref{crestric}) there is equivalent which set of parameters ($c_3,\, c_4$ or $c_1, \, c_2$)  we  preserve, the consideration performed in the same way arrives us to two possibilities:
1) $c_1 =\Lambda, \,   c_2=-\Lambda;$
2) $c_1 =-\Lambda, \,   c_2=\Lambda.$
 Then from the formula (\ref{mumu1}) one gets $\mu\mu_1=\pm1.$ These two different modes may be associated with existence of additional operator of the helicity type. This fact points out on that the wave function substitution (\ref{PsiL}) and the obtained equations (\ref{DireqAsep_L})-(\ref{DireqRsep L}) allow us to obtain the Dirac equation solution without loss of generality.

With notations (\ref{notDH})  the angular system (\ref{DireqAsep_L}) takes the form
\begin{equation}
\begin{split}
\frac{d T_1}{d \theta}+ \Big(\frac{1}{2\tan{\theta}}
-\frac{m}{\sin{\theta}}-2 a \epsilon \tan{\frac{\theta}{2}}
\Big)T_1-
\Lambda T_2=0, \\
\frac{d T_2}{d \theta}+ \Big(\frac{1}{2\tan{\theta}}
+\frac{m}{\sin{\theta}}+2 a \epsilon \tan{\frac{\theta}{2}}
\Big)T_2+ \Lambda T_1=0.
\end{split}
\label{eqangT}
\end{equation}

At vanishing NUT parameter $a=0$, we obtain the case of   Schwarzschild spacetime:
\begin{equation}
\begin{split}
   \Big( \sqrt{\Delta }
\frac{\partial}{\partial r} -
\frac{i \epsilon \rho^2}{\sqrt{\Delta}} \Big) R_3   + i M r R_1 = \Lambda R_4 ,
\\
   \Big( \sqrt{\Delta }
\frac{\partial}{\partial r} -
\frac{i \epsilon \rho^2}{\sqrt{\Delta}}\Big) R_2  + i M  r R_4
= \Lambda R_1 ,
\\
    \Big( \sqrt{\Delta }
\frac{\partial}{\partial r} +
\frac{i \epsilon \rho^2}{\sqrt{\Delta}}\Big) R_1   - i M r R_3
= \Lambda R_2,
\\
   \Big( \sqrt{\Delta }
\frac{\partial}{\partial r} +
\frac{i \epsilon \rho^2}{\sqrt{\Delta}}\Big) R_4   - i M r R_2
= \Lambda R_3;
\end{split}
\label{DireqRsep Schw}
\end{equation}
 here  two possibilities exist:
 $$
I,\;\;
R_4=+ R_1,\quad R_3= + R_2
$$
$$
 \Big(\sqrt{\Delta }
\frac{\partial}{\partial r} +
\frac{i \epsilon \rho^2}{\sqrt{\Delta}} \Big) R_1 - iMr R_2 = \Lambda R_2 , \quad
 \Big(\sqrt{\Delta }
\frac{\partial}{\partial r} -
\frac{i \epsilon \rho^2}{\sqrt{\Delta}} \Big)R_2  + iMr R_1 = \Lambda R_1;
$$

$$II,\;\;
R_4=- R_1,\quad R_3= - R_2 ,
$$
$$
 \Big(\sqrt{\Delta }
\frac{\partial}{\partial r} +
\frac{i \epsilon \rho^2}{\sqrt{\Delta}} \Big) R_1 + iMr R_2 = \Lambda R_2 , \quad
 \Big(\sqrt{\Delta }
\frac{\partial}{\partial r} -
\frac{i \epsilon \rho^2}{\sqrt{\Delta}} \Big)R_2  - iMr R_1 = \Lambda R_1.
$$
The similar result for Schwarzschild  spacetime follows from  equations in \cite{Dariescu2021} for
the Kerr-Newman spacetime after performing  the  limiting procedure to  Schwarzschild  spacetime.

For non-vanishing NUT parameter, we can see from (\ref{DireqRsep L}) that
\begin{equation}
R_1=R_3^*, \, R_2=R_4^*. 
\label{symR}
\end{equation}

\section{  Angular  equations }

 Further, we will solve the angular equations (\ref{eqangT})  in order to get the quantization rule for the separation  parameter
 $\Lambda$.
We introduce new functions $F_i=T_i \sqrt{\sin{\theta}}, i=1,2$
and new variable $z=\sin^2 \theta/2$. The equations (\ref{eqangT})
take the form
\begin{equation}
\begin{split}
F_1'+  \frac{m+4 a \epsilon z}{2z(z-1)}F_1 -\frac{\Lambda
}{\sqrt{(1-z)z}}F_2 =0, \quad
F_2' - \frac{m+4 a \epsilon z}{2z(z-1)}F_2 +\frac{\Lambda
}{\sqrt{(1-z)z}}F_1 =0. \end{split} \label{eqangF}
\end{equation}
The corresponding  2nd order equations are
$$
F_1'' +\frac{2z-1}{2 z (z-1)} F_1' -\left[ \frac{m (m - 1)}{4 z^2
(z-1)^2} + \frac{4 a^2 \epsilon^2}{(z-1)^2}  + \frac{  m (4 a
\epsilon +1 )+2 a \epsilon}{2 z (z-1)^2} + \frac{\Lambda^2}{z
(z-1)} \right]F_1=0,$$ \vspace{-0.9cm} \begin{eqnarray} \label{angFta}
\end{eqnarray}
$$ F_2'' + \frac{2z-1}{2 z (z-1)} F_2' - \left[\frac{m (m + 1)}{4 z^2 (z-1)^2} +
\frac{4 a^2 \epsilon^2}{(z-1)^2}   +
\frac{ m (4 a \epsilon - 1 )- 2 a \epsilon}{2 z (z-1)^2} +
\frac{\Lambda^2}{z (z-1)}\right] F_2=0. $$
\vspace{-0.8cm}
\begin{eqnarray} \label{angFtb}
\end{eqnarray}
We apply the following  substitutions
\begin{equation}
F_1= z^{A}(z-1)^{B} G_1, \quad F_2= z^{C}(z-1)^{D} G_2;
\label{FGchange}
\end{equation}
which lead to the hypergeometric-type equations for $G_1, G_2$:
$$
z(1-z) G_1'' + \Big (\frac{1}{2}(1+4A) -(1+2(A+B))z\Big)G_1'
-\Big((A+B)^2- 4 a^2 \epsilon^2 - \Lambda^2\Big)G_1=0,$$
\vspace{-0.8cm} \begin{eqnarray} \label{eqangG11} \end{eqnarray}
$$z(1-z)G_2'' + \Big (\frac{1}{2}(1+4C)-(1+2(C+D))z\Big
)G_2' -\Big((C+D)^2 -4 a^2 \epsilon^2 - \Lambda^2\Big)G_2=0. $$
\vspace{-0.8cm} \begin{eqnarray} \label{eqangG21}
\end{eqnarray}
So that ($ K_1$ and $K_2$  are yet non-fixed coefficients)
\begin{equation}
F_1 = K_1 \; z^{A}(z-1)^{B} \, _2F_1(a_1,b_1,c_1;z), \quad  F_2 = K_2
\;z^{C}(z-1)^{D} \, _2F_1(a_2,b_2,c_2;z),  \label{eqangFsol}
\end{equation}
$$a_1,\,b_1=A+B\pm \sqrt{4a^2 \epsilon^2+\Lambda^2}, \
c_1=\frac{1}{2}(1+4A); $$$$
a_2,\,b_2=C+D \pm \sqrt{4a^2 \epsilon^2+\Lambda^2},
c_2=\frac{1}{2}(1+4C).$$

We search for  finite solutions $F_1,F_2$. To do it, we have studied the
functions near singular points $z=0,
z=1$. The performed analysis gives the following explicit form of the values $A, B, C, D$:

assuming that $ m>0,\epsilon >0, $
\begin{equation}
\begin{split}
A =\frac{m}{2}, \,  C = \frac{m+1}{2}, \,
B=\frac{4a\epsilon +m+1}{2}, \, D =\frac{4a \epsilon + m}{2};
\end{split}
\nonumber
\end{equation}

assuming that $ m<0,\epsilon >0, $
\begin{equation}
    \begin{split}
A  =-\frac{m-1}{2}, \quad C =-\frac{m}{2}, \quad
B= -\frac{4 a
\epsilon + m}{2}, \quad D = -\frac{4 a \epsilon + m-1}{2}.
    \end{split}
    \nonumber
\end{equation}
Then, introducing the indexes "$+$" and "$-$" for the cases of $m>0$ and $m<0$, respectively, the  parameters of hypergeometric functions in the explicit
form are
%
\begin{equation}
\begin{split}
 a_1^\pm=\frac{1}{2} -
\sqrt{4 a^2 \epsilon^2 +\Lambda^2} \pm (2 a \epsilon  + m),
\quad  b_1^\pm=\frac{1}{2} + \sqrt{4 a^2 \epsilon^2 +\Lambda^2} \pm (2 a \epsilon  + m),
\\
c_1^\pm=1\pm(m-1/2), \, c_2^\pm=
c_1^\pm\pm 1, \,
 a_2^\pm= 
a_1^\pm,
\,
 b_2^\pm=
b_1^\pm. \qquad \qquad
\end{split}
\nonumber
\end{equation}

The quantization rule is found in usual way by imposing the condition that the power series for the hypergeometric function has to be terminated, namely, $b$ has to be a nonpositive integer:
\begin{equation}
\begin{split}
b_1^\pm= \frac{1}{2}+ \sqrt{4 a^2
\epsilon^2 +\Lambda^2} \pm (2 a \epsilon + m) =-n \Rightarrow \qquad \qquad \\
\Lambda^2_\pm =  (n +1/2 \pm m) (n + 1/2 \pm (m+4 a \epsilon)
)   = N (N \pm 4 a \epsilon);\end{split} \label{quantcondp}
\end{equation}
note that at $m<0$ the following constraint should be
satisfied $ - 4 a \epsilon - m >0,
 $
whence if follows $ 4 a \epsilon <  -m +n +1/2 = N.$

Now, substituting the solutions (\ref{eqangFsol})
into eqs.  (\ref{eqangF}), we get   the relations for the coefficients $K_1$ and $K_2$:
\begin{equation}
    \begin{split}
\underline{m>0,} \quad K_1=-\frac{i(1+2m)}{2\Lambda_+}K_2 =
 -\frac{i(1+2m)}{2\sqrt{N(N+4a \epsilon)}}K_2; \\
\underline{m<0,} \quad K_2=-\frac{i(1-2m)}{2\Lambda_-}K_1
= -\frac{i(1- 2 m)}{2\sqrt{N(N - 4a \epsilon)}}K_1.
   \end{split} \nonumber
\end{equation}

Correspondingly, for  initial functions $T_1,T_2$
we obtain the following presentations

\begin{small}
\begin{equation}
\begin{split}
\underline{m>0,}  \hspace{13.3cm} \\ T_1 =
\frac{(1+2m)}{4\sqrt{N(N+4a
\epsilon)}} \;  z^{\frac{(2m-1)}{4}} (1-z)^{2a\epsilon+\frac{2m+1}{4}} 
\, _2F_1 (1+4 a
\epsilon  + 2 m+ n, -n, m+1/2;z);
\\
T_2 =  \frac{1}{2}
\;z^{\frac{2m+1}{4}}(1-z)^{2a\epsilon+\frac{(2m-1)}{4}} 
\, _2F_1 (1+4 a \epsilon  + 2 m+ n,
-n, m+3/2;z).
\end{split}
\end{equation}

\begin{equation}
\begin{split}
\underline{m<0,} \hspace{13.0cm} \\ T_1 = \frac{1}{2}
z^{-\frac{2m-1}{4}}(1-z)^{-(2a\epsilon+\frac{(2m+1)}{4})} 
 \, _2F_1 (1-4 a \epsilon - 2 m + n, -n, - m+3/2;z);
\\
T_2 =   \frac{(1- 2 m)}{4\sqrt{N(N - 4a \epsilon)}}\;
z^{-\frac{(2m+1)}{4}} (1-z)^{-(2a\epsilon+\frac{2m-1}{4})} \times
\, _2F_1 (1 -4  a \epsilon - 2 m +n, -n, -m+1/2;z).
\end{split}
\end{equation}
\end{small}

Let us note that the analytical dependence of the angular solutions on the energy through the combination $a \epsilon$
is similar to that for the Maxwell equations in NUT spacetime \cite{Nouri-Zonoz2004}.

The form of the angular components  for  $m>0$ is illustrated in
Fig. \ref{figT}. They demonstrate the evident effects of a non-vanishing NUT-charge.
Compared with the case of vanishing NUT-charge,
the curves are deformed while the topology of curves remains the same. 

\begin{figure}[hbtp]
\begin{center}
\includegraphics[width=6.5cm,height=4.0cm,angle=0]{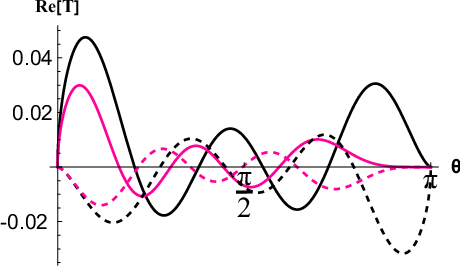}\, \, \,
\includegraphics[width=6.5cm,height=4.5cm,angle=0]{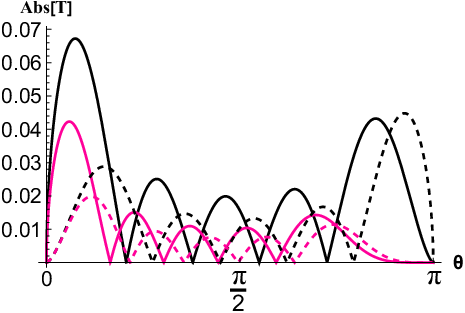}
\end{center}
\caption{The  dependencies of real parts  and absolute values of the functions
$T_1$ (solid lines) and $T_2$ (dashed lines)  on  the variable
$\theta$. Parameters: $m=3/2$; $n=4$; $a=0$ (black), $a=1$ (pink).}
\label{figT}
\end{figure}

\section{ Radial equations}
\subsection{ The case of  massive particle}

We have the known restriction on the
component of the metrical tensor
 $g_{00}>0$, which leads to the inequality   $\Phi>0$, or
$\Delta=r^2-r_g r -a^2>0$. The last leads to the following
physically interpretable region for the radial variable
\begin{equation}
\Delta=r^2-r_g r -a^2=(r-r_1)(r-r_2),\quad
r>r_2 = \frac{1}{2} (r_g+\sqrt{r_g^2+4a^2}),
\nonumber
\end{equation}
where  $ r_2$  determines the
location of the exterior event horizon of the NUT black hole.

It is convenient to introduce the dimensionless quantities,
 $x =
\epsilon r, a \equiv \epsilon a, M \equiv M / \epsilon$.
Then the   system for radial functions
(\ref{Direqseparation a})  transforms to the following one
\begin{equation}
\begin{split}
\Big( \sqrt{\Delta } \frac{d}{d x}-\frac{i a \sqrt{\Delta } }{\rho^2} - \frac{i \rho^2}{\sqrt{\Delta}}\Big) R_3   + i M (x-ia)R_1 =  \Lambda R_4, \\
 \Big( \sqrt{\Delta } \frac{d}{d x}-\frac{i a
\sqrt{\Delta } }{\rho^2} + \frac{i \rho^2}
{\sqrt{\Delta}}\Big) R_4 - i M (x-ia) R_2 = \Lambda R_3 , \\
  \Big( \sqrt{\Delta } \frac{d}{d x}+\frac{i a
\sqrt{\Delta } }{\rho^2} + \frac{i
\rho^2}{\sqrt{\Delta}}\Big) R_1
 - i M (x+ia)R_3  = \Lambda R_2, \\
 \Big( \sqrt{\Delta } \frac{d}{d x}+\frac{i a
\sqrt{\Delta } }{\rho^2} - \frac{i
\rho^2}{\sqrt{\Delta}}\Big) R_2 + i M (x+ia)R_4
 = \Lambda R_1. \end{split}
\label{eqsysRgx}
\end{equation}

Expressing the functions $R_3$ and $R_1$ from the second and
fourth equations in (\ref{eqsysRgx}), respectively, and substituting them into
the first and third equations, we derive the
system of two second order equations relative to functions $R_2$
and $R_4$:
\begin{small}
\begin{equation}
\begin{split}
R_2''+ \Big(\frac{1}{2}\frac{\Delta '}{ \Delta }+\frac{2 i
a}{a^2+x^2}\Big)R_2' + \Big[
-\frac{M^2\left(a^2+x^2\right)}{\Delta
}+\frac{\left(a^2+x^2\right)^2}{\Delta ^2} +\frac{i a \Delta '}{2
\Delta  \left(a^2+x^2\right)} +\frac{i \left(a^2+x^2\right) \Delta
'}{2 \Delta ^2}
\\ -\frac{a^2}{\left(a^2+x^2\right)^2}
-\frac{2 i a x}{\left(a^2+x^2\right)^2} -\frac{\Lambda^2}{\Delta
}-\frac{2 i x}{\Delta }\Big]R_2 + \frac{ i M (x+ia)}{\sqrt{\Delta
} (x-i a)}R_4=0, \qquad \qquad \qquad
\\
R_4''+ \Big(\frac{1}{2}\frac{\Delta '}{ \Delta }-\frac{2 i
a}{a^2+x^2}\Big)R_4' + \Big[ -\frac{M^2
\left(a^2+x^2\right)}{\Delta
}+\frac{\left(a^2+x^2\right)^2}{\Delta ^2} -\frac{i a \Delta '}{2
\Delta  \left(a^2+x^2\right)} -\frac{i \left(a^2+x^2\right) \Delta
'}{2 \Delta ^2}
\\ -\frac{a^2}{\left(a^2+x^2\right)^2}
+\frac{2 i a x}{\left(a^2+x^2\right)^2} -\frac{\Lambda^2}{\Delta
}+\frac{2 i x}{\Delta }\Big]R_4 - \frac{ i M (x-ia)}{\sqrt{\Delta
} (x+i a)}R_2=0; \qquad \qquad \qquad
\end{split}
 \label{eqRMass24}
\end{equation}
\end{small}
By the linear combinations of the equations (\ref{eqsysRgx}) in
the same way, we get the  system for the functions $R_3,$ $R_1$:
\begin{small}
\begin{equation}
\begin{split}
R_1''+ \Big(\frac{1}{2}\frac{\Delta '}{ \Delta }+\frac{2 i
a}{a^2+x^2}\Big)R_1' + \Big[ -\frac{M^2
\left(a^2+x^2\right)}{\Delta
}+\frac{\left(a^2+x^2\right)^2}{\Delta ^2}
+\frac{i a \Delta '}{2
\Delta  \left(a^2+x^2\right)}
-\frac{i \left(a^2+x^2\right) \Delta '}{2 \Delta ^2}
\\ -\frac{a^2}{\left(a^2+x^2\right)^2}
-\frac{2 i a
x}{\left(a^2+x^2\right)^2}-\frac{\Lambda^2}{\Delta }+\frac{2 i
x}{\Delta }\Big]R_1 - \frac{ i M (x+ia)}{\sqrt{\Delta } (x-i
a)}R_3=0, \qquad \qquad \qquad
\\
R_3''+ \Big(\frac{1}{2}\frac{\Delta '}{ \Delta }-\frac{2 i
a}{a^2+x^2}\Big)R_3' + \Big[ -\frac{M^2
\left(a^2+x^2\right)}{\Delta
}+\frac{\left(a^2+x^2\right)^2}{\Delta ^2}
-\frac{i a \Delta '}{2
\Delta  \left(a^2+x^2\right)}
+\frac{i \left(a^2+x^2\right) \Delta '}{2 \Delta
^2} \\ -\frac{a^2}{\left(a^2+x^2\right)^2}+\frac{2 i a
x}{\left(a^2+x^2\right)^2}
-\frac{\Lambda^2}{\Delta }-\frac{2 i
x}{\Delta }\Big]R_3 + \frac{ i M (x-ia)}{\sqrt{\Delta } (x+i
a)}R_1=0. \qquad\qquad\qquad
\end{split}
 \label{eqRMass13}
\end{equation}
\end{small}

Taking into account the symmetry (\ref{symR}), it is enough to consider the subsystem
for variables $R_1, R_3$.
In order to simplify the problem, we make substitutions
\begin{equation}
Z_1= V_1 R_1, \qquad Z_3= V_3 R_3, \label{ZamenaV}
\end{equation}
where

\begin{equation}
V_1=\frac{ \sqrt{x-i a} }{\sqrt{x+i
a}}\left(x-x_1\right)^{-\frac{1}{2}- i x_1}
\left(x-x_2\right){}^{-\frac{1}{2}-i x_2} e^{-i
\left(x-x_1\right)}, \label{V1}
\end{equation}
\begin{equation}
V_3=\frac{i  \sqrt{x+i a} }{\sqrt{x-i a}}\left(x-x_1\right){}^{- i
x_1} \left(x-x_2\right){}^{-i x_2}e^{-i x }. \label{V3}
\end{equation}
This results in
$$Z_1''+\left(\frac{3+4 i x_1}{2
\left(x-x_1\right)}+\frac{i \left(4 x_2-3 i\right)}{2
\left(x-x_2\right)}+2 i\right) Z_1'
$$$$-\frac{ \left(M^2 \left(a^2+x^2\right)+\Lambda^2-4 i
x-1\right)}{\left(x-x_1\right) \left(x-x_2\right)}Z_1-\frac{M
}{\left(x-x_1\right) \left(x-x_2\right)}Z_3=0,
$$
$$
Z_3'' +\left(\frac{1+4 i x_1}{2 \left(x-x_1\right)}+\frac{i
\left(4 x_2-i\right)}{2 \left(x-x_2\right)}+2 i\right) Z_3'
-\frac{ \left(M^2 \left(a^2+x^2\right)+\Lambda^2\right)}
{\left(x-x_1\right) \left(x-x_2\right)}Z_3 -M Z_1 =0.
$$
Eliminating the variable  $Z_1$, we derive the fourth order
equation for  $Z_3$:
\begin{small}
$$
Z_3^{(4)}+ 4 i \left(1+\frac{2 x_1-i}{2(x-x_1)}+\frac{2
x_2-i}{2(x-x_2)}\right) Z_3^{(3)}-\Big(4 +2 M^2+\frac{1+16
x_1^2}{4\left(x-x_1\right)^2}+\frac{1+16 x_2^2}{
4\left(x-x_2\right){}^2}
   $$
$$
+\frac{4+4\left(\Lambda^2+M^2(a^2+x_1^2)\right)+\left(3i-4x_1\right)^2}{2
\left(x-x_1\right) \left(x_1-x_2\right)}
-\frac{4+4\left(\Lambda^2+M^2(a^2+x_2^2)\right)+\left(3i-4x_2\right)^2}{2
\left(x-x_2\right) \left(x_1-x_2\right)}\Big)Z_3^{(2)}
$$
$$
-\Big(4 i M^2-\frac{1+16 x_1^2}{4\left(x-x_1\right)^3}-\frac{1+16
x_2^2}{
4\left(x-x_2\right){}^3}+\frac{1+16ix_1\left(\Lambda^2+M^2(a^2+x_1^2)-ix_1\right)}{4(x-x_1)^2(x_1-x_2)}
   $$
   $$
-\frac{1+16ix_2\left(\Lambda^2+M^2(a^2+x_2^2)-ix_2\right)}{4(x-x_2)^2(x_1-x_2)}
+\frac{4(1+2ix_1)(M^2x_1-i)}{(x-x_1)(x_1-x_2)}
-\frac{4(1+2ix_2)(M^2x_2-i)}{(x-x_2)(x_1-x_2)}\Big)Z_3^{(1)}
$$
$$+\Big(M^4+\frac{i(i+4x_1)\left(\Lambda^2+M^2(a^2+x_1^2)\right)}{2(x-x_1)^3(x_1-x_2)} -\frac{i(i+4x_2)\left(\Lambda^2+M^2(a^2+x_2^2)\right)}{2(x-x_2)^3(x_1-x_2)}
   $$
$$
+\frac{\left(\Lambda^2+M^2(a^2+x_1^2)\right)\left(1-4ix_1+2\left(\Lambda^2+M^2(a^2+x_1^2)\right)\right)}{2(x-x_1)^2(x_1-x_2)^2}
$$$$+\frac{\left(\Lambda^2+M^2(a^2+x_2^2)\right)\left(1-4ix_2+2\left(\Lambda^2+M^2(a^2+x_2^2)\right)\right)}{2(x-x_2)^2(x_1-x_2)^2}
   $$
   $$-\frac{1}{2(x-x_1)(x_1-x_2)^3}\Big(\left(1-4ix_2\right)\left(\Lambda^2+M^2(a^2+x_1^2)\right)+\left(1-4ix_1\right)\left(\Lambda^2+M^2(a^2+x_2^2)\right)
   $$$$+4\left(\Lambda^4-M^4(a^2+x_1^2)^2\right)
   +4M^2(1+2ix_1)(x_1-x_2)^2\Big)
   $$
   $$
+\frac{1}{2(x-x_2)(x_1-x_2)^3}\Big(\left(1-4ix_2\right)\left(\Lambda^2+M^2(a^2+x_1^2)\right)+\left(1-4ix_1\right)\left(\Lambda^2+M^2(a^2+x_2^2)\right)
$$$$+4\left(\Lambda^4-M^4(a^2+x_2^2)^2\right)
   +4M^2(1+2ix_2)(x_1-x_2)^2\Big)\Big)Z_3=0.
   $$
\end{small}

In order to study the radiation emitted by the black hole, let
us find the appro\-xi\-mate equation for the radial function near the
exterior horizon $x \rightarrow x_2$. Preserving only the largest terms in the previous equation, we obtain
\begin{equation}
\begin{split}
Z_3{}^{(4)}+\frac{2 i \left(2 x_2-i\right)
}{x-x_2}Z_3{}^{(3)}-\frac{\left(16 x_2^2+1\right)
}{4 \left(x-x_2\right){}^2}Z_3''  \qquad \\ +\frac{\left(16 x_2^2+1\right) }{4
\left(x-x_2\right){}^3}Z_3'
+\frac{i \left(4 x_2+i\right) \left(M^2
\left(a^2+x_2^2\right)+\Lambda ^2\right)}{2 \left(x-x_2\right){}^3
\left(x_2-x_1\right)}Z_3 =0.
\end{split}
\label{eqZ3e4x2}
\end{equation}
The possible structure for the solutions is $(x-x_2)^A$. Substituting this into  eq. (\ref{eqZ3e4x2})
and neglecting the last term, we get
$$
(-2 + A) A (-3 + 2 A + 4 i x_2) (-1 + 2 A + 4 i x_2) = 0;
$$
whence it follows
$A=0, \, \, A=2, \, \, A=\frac{1}{2}  (1 - 4 i x_2), \, \, A=\frac{1}{2} (3  - 4 i x_2).$
Taking into consideration the form of  $V_3$ and the expressions for multipliers when separating the variables (\ref{separ1}),
 we obtain the behavior of  radial component near $x_2$:
$$\Delta^{-\frac{1}{4}}\rho_{-}^{-\frac{1}{2}}R_3\sim (x-x_2)^{i x_2 -\frac{1}{4}}; \,\, (x-x_2)^{i x_2 +\frac{7}{4}}; \,\, (x-x_2)^{\frac{1}{4} - i x_2 };
\,\,(x-x_2)^{\frac{5}{4} - i x_2 }.$$
The first two solutions correspond to the outgoing waves and the
last two solutions determine the ingoing waves. The first solution possess the singularity
at $x=x_2$, because of that it is nonphysical one and is rejected from the following consideration.
Taking into account the  factor $e^{i \epsilon t}$,  the wave solutions are (recall that $x=\epsilon r$)
$$\Psi_{in} = e^{-i \epsilon t} (r-r_2)^{-i \epsilon r_2 +\frac{1}{4}}, \, \, e^{-i \epsilon t} (r-r_2)^{-i \epsilon r_2 +\frac{5}{4}},$$
$$\Psi_{out}(r>r_2) = e^{-i \epsilon t} (r-r_2)^{+i \epsilon r_2 +\frac{7}{4}},$$
or in ordinary notations
$$\Psi_{in} =e^{-i \omega t} (r-r_2)^\gamma (r-r_2)^{-\frac{i}{2 \kappa_+}(\omega-\omega_0)},$$
$$\Psi_{out}(r>r_2) =e^{-i \omega t} (r-r_2)^{\gamma} (r-r_2)^{\frac{i}{2 \kappa_+}(\omega-\omega_0)},$$
where $\omega\equiv \epsilon$, $\kappa_+= 1/2r_2$, $\omega_0=0$,
$\gamma$ is the attenuation coefficient.

The tortoise and Eddington-Finkelstein coordinates are defined by
\cite{Sannan1988,Vieira2015}
$$\ln(r-r_2)=\frac{1}{r_2^2+a^2} \left. \frac{\Delta}{dr} \right|_{r=r_2} r^* =
\frac{1}{r_2(r_2-r_1)} \left. \frac{(r-r_1)(r-r_2)}{dr} \right|_{r=r_2} r^*=\frac{1}{r_2} r^* =2 \kappa_+ r^*,$$
$$v=t+r^*, \quad \Rightarrow \quad t=v- \frac{1}{2 \kappa_+} \ln(r-r_2).$$
In new coordinates, the solutions take the form
$$\Psi_{in} =e^{-i \omega v} (r-r_2)^{\gamma},$$
$$\Psi_{out}(r>r_2) =e^{-i \omega v} (r-r_2)^{\gamma} (r-r_2)^{\frac{i}{\kappa_+} \omega}.$$
$\Psi_{in}$ is well-behaved when analytically extended inside the horizon $x<x_2$.
$(r-r_2)^{\gamma}$ is also well-behaved function while $(r-r_2)^{\frac{i}{\kappa_+} \omega}$
has a logarithmic singularity at $x=x_2$. The  unique continuation of the function according to the Damour-Ruffini method
is rotating $-\pi$ through the lower-half complex $r$ plane.
Then, the analytic continuation of a real damped part of the outgoing wave on the horizon surface $r<r_2$ is
$$\Psi_{out}(r<r_2) =e^{-i \omega v} (r-r_2)^{\gamma} (r_2-r)^{\frac{i}{\kappa_+} \omega}e^{\frac{\pi}{\kappa_+}\omega}.$$
According to \cite{Sannan1988,Dariescu2021,Vieira2015}, a  scattering probability
\begin{equation}
\Gamma=\left|{\frac{\Psi_{out}(x>x_2)}{\Psi_{out}(x<x_2)}}\right|^2
\label{scatprob}
\end{equation}
is the probability of creating a particle-antiparticle pair just outside the exterior
horizon. Substituting the outgoing wave into the formula
(\ref{scatprob}), we get
\begin{equation}
\Gamma=e^{-4\pi x_2}=e^{-4\pi\epsilon r_2}.
\end{equation}
The mean number $\bar{N}_\epsilon$ of fermions emitted with a
given energy (in a fixed mode) is determined by the relation (ignoring  the  backscattering effect):
\begin{equation}
\bar{N}_\epsilon=\frac{\Gamma}{\Gamma+1}=\frac{1}{1+e^{4\pi\epsilon r_2}}.
\end{equation}
We get the Fermi-Dirac distribution
\begin{equation}
\bar{N}_\epsilon=\frac{1}{1+e^{(\epsilon-\epsilon_0)/T}}, \qquad  T=\frac{1}{4 \pi r_2}=\frac{1}{2 \pi (r_g+ \sqrt{r_g^2+ 4 a^2})}.
\end{equation}
This expression  for Hawking  temperature coincides with the result obtained  for the   Taub-NUT black hole in
\cite{Liu2022}.

As $r_2>r_g$ at all real
values $a$, the Hawking temperature decreases with the increase of the
NUT charge, this corresponds to decreasing the probability of
particle-antiparticle pair production.
It should be noted that in contrast to the  Kerr-Newman spacetime
\cite{Dariescu2021}, the obtained scattering probability does not depend on the third projection of the total angular momentum  $m$.

The asymptotic behavior at infinity for the function $Z_3$ can be represented as the superposition of ingoing and outgoing damped waves (for detail see Supplement C)
\begin{equation}
\begin{split}
 \Delta^{-1/4}\rho_{-}^{-1/2}R_3 \sim c_1 R_{\text{out}} + c_2 R_{\text{in}} \\
 \sim c_1  x^{ -\frac{3}{2} + i x_g \Big(1-\frac{\Gamma^2 }{2 \sqrt{1-M^2}}\Big)
\pm \frac{1}{2} \sqrt{1- \frac{  12 \Gamma^2 +2 \Gamma(8+ \Gamma^2 -9 i\Gamma) x_g +\Gamma ^4 x_g^2 }{1-M^2}}}
 e^{ + i \sqrt{1-M^2} x} \\ + c_2  x^{ -\frac{3}{2} + i x_g \Big(1-\frac{\Gamma^2 }{2 \sqrt{1-M^2}}\Big)
\pm \frac{1}{2} \sqrt{1- \frac{  12 \Gamma^2 +2 \Gamma(8+ \Gamma^2 -9 i\Gamma) x_g +\Gamma ^4 x_g^2 }{1-M^2}}}
 e^{ - i \sqrt{1-M^2} x};
 \end{split}
\end{equation}
where $\Gamma = -i\Big(1 \pm \sqrt{1-M^2}\Big)$. Let us recall, that here $M$ is renormalized on the energy $\epsilon$ ($M \equiv M/\epsilon$).
As should be expected, the asymptotic behavior at infinity does not depend on the NUT parameter, but is determined  by the mass of the black hole $M_{BH}$ ($x_g=2M_{BH}$), and the mass and the energy of the particle.

Let us show that there exists specific peculiarity  due to the non-vanishing  NUT-charge.
Indeed, taking into account the identities
$R_1=R_3^*$ and  $R_2=R_4^*$, let us perform the following  combination over equations
in (\ref{eqsysRgx}):
$$
eq.1\times R_1 + eq.3 \times R_3 - eq.2 \times R_2 - eq. 4 \times R_4,
$$
then we arrive at
\begin{equation}
\begin{split}
\sqrt{\Delta}(R_1 R_1^*-R_2 R_2^*)'
+ i M \Big[(x- ia) \Big(R_1^2 + R_2 ^2\Big)  - (x+ ia) \Big( \Big(R_1^*\Big)^2 + \Big(R_2^*\Big) ^2\Big) \Big]=0, \end{split}
\label{AbsR1R2}
\end{equation}
here the  derivative over $r$ is denoted by a prime.

For  Schwarzschild metric, at  $a=0$, from the system of four equations (\ref{eqsysRgx})
we have the only two independent and conjugate equations, $R_1=R_2^*$, and correspondingly the second term in eq.
(\ref{AbsR1R2}) vanishes identically. Therefore,  the absolute values of radial components are equal, $|R_1|=|R_2|$.
As one can see from (\ref{AbsR1R2}), NUT charge  
results in non-equal  amplitudes, $|R_1|\neq|R_2|$,
for massive particles (with $M\neq0$). 
The situation for massless case seems to be completely different, as  $M=0$ the equation (\ref{AbsR1R2}) takes the form: $(R_1 R_1^*-R_2 R_2^*)'=0$.
The solution of the last gives
the amplitude equality $|R_1|=|R_2|$. 

\subsection{ The case of  massless particle}

In the massless case,  the equations (\ref{eqsysRgx}) are simplified: they may be
divided into two unlinked subsystems
\begin{equation}
\begin{split}
\Big( \sqrt{\Delta } \frac{d}{d x}-\frac{i a \sqrt{\Delta } }{\rho
^2} - \frac{i
\rho^2}{\sqrt{\Delta}}\Big) R_3   =  \Lambda R_4, \\
 \Big( \sqrt{\Delta } \frac{d}{d x}-\frac{i a
\sqrt{\Delta } }{\rho ^2} + \frac{i \rho^2}{\sqrt{\Delta}}\Big)
R_4  = \Lambda R_3;
\end{split}
\label{eqsysRgM034}
\end{equation}
\begin{equation}
\begin{split}
  \Big( \sqrt{\Delta } \frac{d}{d x}+\frac{i a
\sqrt{\Delta } }{\rho ^2} + \frac{i
\rho^2}{\sqrt{\Delta}}\Big) R_1   = \Lambda R_2, \\
 \Big( \sqrt{\Delta } \frac{d}{d x}+\frac{i a
\sqrt{\Delta } }{\rho ^2} - \frac{i \rho^2}{\sqrt{\Delta}}\Big)
R_2
 = \Lambda R_1.
\end{split}
\label{eqsysRgM012}
\end{equation}
The system (\ref{eqsysRgM034}) is conjugated to (\ref{eqsysRgM012}),
by this reason  it is enough to consider only the subsystem (\ref{eqsysRgM012}).
We readily find the 2nd order equations for separate functions
\begin{equation}
\begin{split}
\Delta R_1 '' + \Big[\frac{2 i a \Delta}{a^2+x^2}
+\frac{1}{2}\Delta' \Big] R_1'  +\Big[ -\Lambda^2+ 2 i x + \frac{
(a^2+x^2)^2}{\Delta} \\ - \frac{a (a + 2 i x )
\Delta}{(a^2+x^2)^2} +\frac{i a \Delta'}{2(a^2+x^2)} - \frac{i
 (a^2+x^2) \Delta'}{2 \Delta}\Big] R_1=0,
\end{split}
\label{eqR1x}
\end{equation}
\begin{equation}
\begin{split}
\Delta R_2 '' + \Big[\frac{2 i a \Delta}{a^2+x^2}
+\frac{1}{2}\Delta' \Big] R_2'  +\Big[ -\Lambda^2- 2 i x + \frac{
(a^2+x^2)^2}{\Delta} \\ - \frac{a (a + 2 i x )
\Delta}{(a^2+x^2)^2} +\frac{i a \Delta'}{2(a^2+x^2)} + \frac{i
(a^2+x^2) \Delta'}{2 \Delta}\Big] R_2=0.
\end{split}
\label{eqR2x}
\end{equation}

Applying the substitution (\ref{ZamenaV})--(\ref{V1}), from eq.
(\ref{eqR1x}) we obtain the following equation for $Z_1$:
\begin{equation}
\begin{split}
Z_1''+\Big(\frac{i (4 x_1-3 i)}{2 (x-x_1)}+\frac{i (4x_2-3 i)}{2
(x-x_2)}+2 i\Big)Z_1'   + \frac{1+ 4 i x - \Lambda^2}{(x-x_1)
(x-x_2)}Z_1=0.  \label{eqM0Z1}
\end{split}
\end{equation}
In the new variable
$$
v=\frac{x-x_2}{x_1-x_2},
$$
eq. (\ref{eqM0Z1})
takes the form
\begin{equation}
\begin{split}
Z_1'' + \Big( (z_1-z_2)+\frac{z_2}{v} + \frac{z_1}{v-1} \Big)
Z_1'
+  \frac{-2-\Lambda^2 +  2 z_2 + 2 (z_1-z_2) v }{ v (v-1)}Z_1
=0,
\end{split}
\label{eqZ1v}
\end{equation}
where $z_1=2ix_1+3/2$, $z_2=2ix_2+3/2$.
Its general  solution  can be expressed in terms of
the confluent  Heun functions:
$$
Z_1= C_{11} \mbox{HeunC} \Big[ q_{11}; \alpha_{11}, \gamma_{11}, \delta_{1},
\varepsilon;  v \Big]
+ C_{12} v^{1-\gamma_{11}} \mbox{HeunC} \Big[ q_{12}; \alpha_{12}, \gamma_{12}, \delta_{1},
\varepsilon;  v \Big],
$$
 where
 \begin{equation}
 \begin{split}
q_{11} = 2+\Lambda^2 -  2 z_2, \quad \alpha_{11} = 2 (z_1-z_2), \quad \varepsilon = (z_1-z_2), \quad
 \gamma_{11} = z_2, \quad \delta_{1} = z_1;  \\ q_{12} = (-\delta_1+\varepsilon)(1-\gamma_{11})q_{11}, \quad \alpha_{12} = \alpha_{11}+\varepsilon(1-\gamma_{11}),
\quad \gamma_{12} = 2-\gamma_{11}. \quad \end{split} \nonumber
 \end{equation}

 Let us turn back to the original variable $R_1$:
\begin{equation}
\begin{split}
R_1  \sim  e^{i \epsilon \left(r-r_1\right)}
\left(r-r_1\right)^{\frac{1}{2}+i \epsilon r_1}
(r-r_2)^{\frac{1}{2}+i \epsilon r_2} \frac{ \sqrt{r+i
a}}{\sqrt{r-i a}}  \qquad \qquad \qquad \qquad \\
\times \Big (C_{11} \mbox{HeunC} (  q_{11}; \alpha_{11}, \gamma_{11}, \delta_{1},
\varepsilon;  \frac{r-r_2}{r_1-r_2} )
+ C_{12}
v^{1-\gamma_{11}} \mbox{HeunC} ( q_{12}; \alpha_{12}, \gamma_{12}, \delta_{1},
\varepsilon;   \frac{r-r_2}{r_1-r_2})\Big ).
\end{split}
\label{solR1M0}
\end{equation}
In turn,    with the use of substitution
 $Z_2 = V_2 R_2,$ where
$$
V_2= \frac{e^{-i \left(x-x_1\right)} \sqrt{a+i x}
\left(-x+x_1\right)^{-i x_1} (-x+x_2)^{-i x_2}}{\sqrt{a-i x}};
$$
for the function $R_2$ related to  eq. (\ref{eqR2x}) we derive
\begin{equation}
\begin{split}
R_2  \sim  e^{i \epsilon \left(r-r_1\right)}
\left(r-r_1\right)^{i \epsilon r_1}
(r-r_2)^{i \epsilon r_2} \frac{ \sqrt{r+i
a}}{\sqrt{r-i a}}  \qquad \qquad \qquad \qquad \\
\times \Big (C_{21} \mbox{HeunC} (  q_{21}; \alpha_{21}, \gamma_{21}, \delta_{2},
\varepsilon;  \frac{r-r_2}{r_1-r_2} )
 + C_{22}
v^{1-\gamma_{21}} \mbox{HeunC} (q_{22}; \alpha_{22}, \gamma_{22}, \delta_{2},
\varepsilon;   \frac{r-r_2}{r_1-r_2}) \Big);
\end{split}
\label{solR2M0}
\end{equation}
\begin{equation}
\begin{split} q_{21} = -\Lambda^2, \quad \alpha_{21} = 0, \quad \gamma_{21} = z_2-1, \quad \delta_{2} = z_1-1;  \qquad \qquad \\
 q_{22} = (-\delta_2+\varepsilon)(1-\gamma_{21})q_{21}, \quad \alpha_{22} = \alpha_{21}+\varepsilon(1-\gamma_{21}),
\quad \gamma_{22} = 2-\gamma_{21}. \end{split} \nonumber
\end{equation}

\begin{figure}[hbtp]
\begin{center}
\includegraphics[width=7.0cm,height=4.5cm,angle=0]{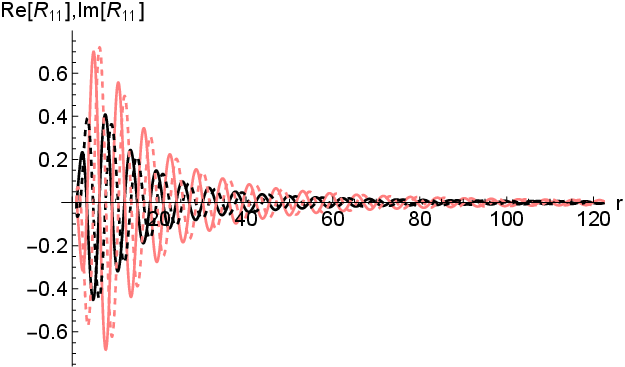}\, \, \,
\includegraphics[width=7.0cm,height=4.5cm,angle=0]{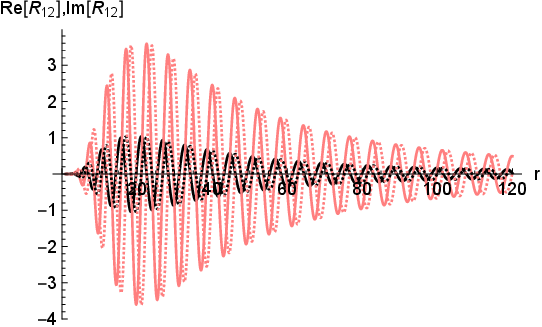}
\end{center}
\caption{The  real (solid lines) and imaginary (dashed lines) parts  of the functions $R_{11}$, $R_{12}$ of the wave function (\ref{tofig2}) for the case of massless fermion. Parameters: $r_g=1$; $\epsilon=1$;$N=3$; $a=0$ (black), $a=0.1$ (pink).}
\label{figR}
\end{figure}

Taking into account the multipliers evolved at separation procedure, the complete radial function (for example, for the component $R_1$) can be presented as follows
\begin{equation}
\Delta^{-1/4}\rho_{+}^{-1/2}R_{1}=C_{11}R_{11}+C_{12}R_{12}.
\label{tofig2}
\end{equation}
The behavior of the
this functions $R_{11}$ and $R_{12}$ is illustrated in Fig. \ref{figR}

Series expansion of the confluent Heun function 
$$
Z(v)=\mbox{HeunC} \Big[  q; \alpha, \gamma, \delta,
\varepsilon;  v \Big]=v(v-1) \sum_{n=0}^\infty c_n v^n
$$
around  the regular singular point $v=0$ ($x=x_2$) gives
the three-term recurrence relation
\begin{equation}
\begin{split}
n \geq 2, \,\, C_{n+1} c_{n-1} + B_{n+1} c_n  - A_{n+1} c_{n+1} =0, \hspace{3.0cm} \\ A_{n+1}=(n+1)(\gamma +n), \quad B_{n+1}= (n (\gamma +\delta + n -\varepsilon -1) -q),  \quad
C_{n+1}= (\alpha +(n-1) \varepsilon ).
\end{split}
\nonumber
\end{equation}
Let us restrict ourselves to the transcendental confluent Heun functions, which are obtained by imposing
the $\delta_N$ -condition  \cite{Fiziev2010,Slavyanov2000}:
$C_{n+2} = 0$, whence it follows
$\alpha + n \varepsilon =0.$
Only the  components with $C_{12}$ and $C_{22}$ in general solutions
(\ref{solR1M0}), (\ref{solR2M0}) satisfy this constrain, at this we obtain the imaginary energies
$$\epsilon_I = -i\frac{3+2n}{4  r_2}=-i\frac{3+2n}{2  (r_g+\sqrt{r_g^2+4a^2})}.$$
The derived energy quasispectrum determines the frequencies
which represent the scattering
resonances of the fields in the black hole spacetime  \cite{Vieira2016,Dariescu2021}.
The resonances characterize the poles of the  transmission (re\-flec\-tion) amplitudes dependencies on the energy.
In opposite to the case of Kerr black hole \cite{Dariescu2021}, the values of resonant energies do not possess the real part and are imaginary ones, similarly to the Schwarzschild case. It seems to be related with  absence of the bound states. Nevertheless, the non-vanishing NUT charge influences the wave characteristics. Thus, the resonant energies associated with the massless fermion
are decreased with the rise of NUT charge $a$  compared with the Schwarzschild black hole levels.

\subsubsection{Effective potential }
Let us discuss the possibility of describing the system under consideration using the concept of the effective potential.
To this end, we introduce the new variables $f(x)=R_1+R_2, g(x)=i(R_1-R_2)$. Then the radial equations
(\ref{eqsysRgM012}) take the form 
\begin{equation}
\begin{split}
\sqrt{\Delta} f'+ \Big(\frac{i a \sqrt{\Delta }}{a^2+x^2}-\Lambda \Big)f + \frac{ \left(a^2+x^2\right)}{\sqrt{\Delta }}g =0, \nonumber \\
\sqrt{\Delta} g' + \Big(\frac{i a \sqrt{\Delta }}{a^2+x^2} +
\Lambda \Big)g - \frac{ \left(a^2+x^2\right)}{\sqrt{\Delta }}f =0.
\end{split}
\end{equation}
The corresponding second order equations are
\begin{equation}
\begin{split}
f''+ \left(\frac{\Delta '}{\Delta }-\frac{2}{x+i a}\right)f'
+\left(\frac{\left(a^2+x^2\right)^2}{\Delta ^2}+\frac{i a \Delta
'}{\Delta  \left(a^2+x^2\right)} \right.  \\
\left. +\frac{2 \Lambda  x}{\sqrt{\Delta
}\left(a^2+x^2\right)}-\frac{a (a+4 i x)}{\left(a^2+x^2\right)^2}
-\frac{\Lambda  \Delta '}{2 \Delta ^{3/2}}-\frac{\Lambda ^2}{\Delta }\right)f =0, \end{split} \label{eqf2} \end{equation}
\begin{equation}
\begin{split}
g''+ \left(\frac{\Delta '}{\Delta }-\frac{2}{x+i a}\right)g'
+\left(\frac{\left(a^2+x^2\right)^2}{\Delta ^2}+\frac{i a \Delta
'}{\Delta  \left(a^2+x^2\right)}  \right. \\
\left.-\frac{2 \Lambda  x}{\sqrt{\Delta
}\left(a^2+x^2\right)}-\frac{a (a+4 i x)}{\left(a^2+x^2\right)^2}
+\frac{\Lambda  \Delta '}{2 \Delta ^{3/2}}-\frac{\Lambda
^2}{\Delta }\right)g =0.  \end{split} \label{eqg2}
\end{equation}

Let us find a special variable $w$, which generalizes tortoise-like coordinate and transforms the equations
(\ref{eqf2})--(\ref{eqg2}) to the structure
\begin{equation}
\Big[\frac{d^2}{dw^2}+P(w)\Big]f=0, \qquad
\Big[\frac{d^2}{dw^2}+Q(w)\Big]g=0, \label{eqfg2w}
\end{equation}
where  $P(w)$ and $Q(w)$ may be considered as effective
squared linear momentums, shortly -- potentials. The
variable $w$ is determined by the equation
\begin{equation}
\frac{d^2 w}{dx^2} + \Big[\frac{\Delta '}{\Delta }-\frac{2}{x+i a}
\Big] \frac{d w}{dx}=0,
\end{equation}
whence we readily find
\begin{equation}
\begin{split}
w= x+\frac{\left(a-i x_1\right)^2 \ln \left(x-x_1\right)-\left(a-i x_2\right)^2 \ln \left(x-x_2\right)}{x_2-x_1},
\quad x>x_2>x_1.
\end{split}
\nonumber
\end{equation}

\begin{figure}[hbt]
\begin{center}
(a), \hspace{4.0cm} (b), \hspace{4.0cm}(c)  \\
\includegraphics[width=4.5cm,height=3.0cm,angle=0]{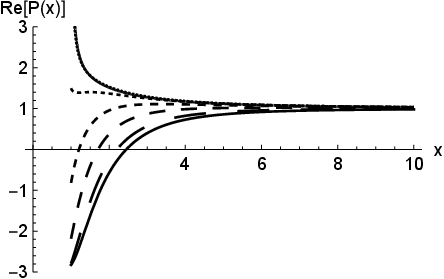} \,
 \includegraphics[width=4.5cm,height=3.0cm,angle=0]{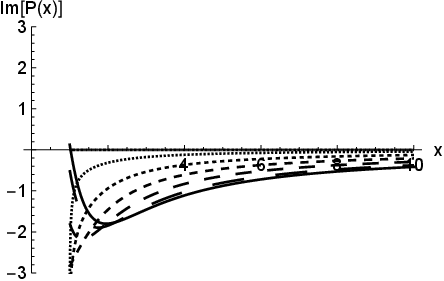}  \,
  \includegraphics[width=4.5cm,height=3.0cm,angle=0]{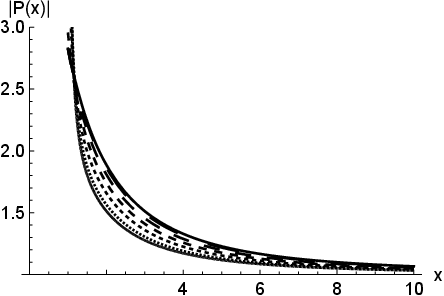}
\end{center}
\caption{The  real (a) and imaginary (b) parts  and absolute values (c) of the potential
$P(x)$ in dependence on the NUT parameter $a$. Values of $a$ decreased  sequentially ($1$, $0.9$, $0.7$,
$0.5$, $0.3$, $0.1$, $0$) from solid to pointed lines. }
\label{figP}
\end{figure}

In the limiting case of the Schwarzschild metric  ($a=0$) with
horizon $x_g=1$, one gets $w= x+ \ln(x-1)$ which is ordinary tortoise coordinate, in this case eqs.
(\ref{eqf2})--(\ref{eqg2}) coincides with the equations obtained in \cite{5}.

The potentials $P$ and $Q$ are
\begin{equation}
\begin{split}
P=\frac{\left(x-x_1\right){}^2 \left(x-x_2\right){}^2}{(a-i x)^4}
\left(\frac{\left(a^2+x^2\right)^2}{\Delta ^2}+\frac{i a \Delta
'}{\Delta  \left(a^2+x^2\right)} \right.  \\
\left. +\frac{2 \Lambda x}{\sqrt{\Delta
}\left(a^2+x^2\right)}-\frac{a (a+4 i x)}{\left(a^2+x^2\right)^2}
-\frac{\Lambda  \Delta '}{2 \Delta ^{3/2}}-\frac{\Lambda ^2}{\Delta }\right),  \\
Q=\frac{\left(x-x_1\right){}^2 \left(x-x_2\right){}^2}{(a-i x)^4}
\left(\frac{\left(a^2+x^2\right)^2}{\Delta ^2}+\frac{i a \Delta
'}{\Delta  \left(a^2+x^2\right)} \right.   \\
\left.-\frac{2 \Lambda x}{\sqrt{\Delta
}\left(a^2+x^2\right)}-\frac{a (a+4 i x)}{\left(a^2+x^2\right)^2}
+\frac{\Lambda  \Delta '}{2 \Delta ^{3/2}}-\frac{\Lambda
^2}{\Delta }\right). \end{split}
\end{equation}
The potential $P(x)$ is illustrated in Fig. \ref{figP}.
As one can see from Fig. \ref{figP}(a)-(b), when the NUT-charge increases, the real and imaginary parts change the character of their behavior
to the opposite.
The dependence of the absolute value of the effective potentials $P(x)$ and $Q(x)$ is presented by the monotonically decreasing curve, similarly to the case of the Schwarzschild metric, which evidences the
absence of the bound states for such systems. However, it should be specially noted that
the potentials are complex-valued functions, while at imaginary values of NUT-charge $a = i|a|$ they become real-valued.
At the exterior horizon $x \rightarrow x_2$ they behave as
\begin{small}
$$P=Q=\frac{(x_2  - i a)^2}{(x_2 + i a)^2}
=\frac{\left(a^2-x_2^2\right){}^2-4 a^2
x_2^2}{\left(a^2+x_2^2\right){}^2}+i\frac{4  a x_2
\left(a^2-x_2^2\right)}{\left(a^2+x_2^2\right){}^2}$$
\end{small}
and at infinity $x \rightarrow \infty$, the potentials tend to the unit, $P=Q=1.$

\section{ Black hole with a small  NUT parameter}

In the limiting case of the Schwarzschild black hole,  the radial components of the wave function
obey the conditions
\begin{equation}
R_4=-R_1, \qquad R_3=-R_2,
\label{R34cond}
\end{equation}
 as one can see from eq. (\ref{eqsysRgx}).
Let us assume small value of the NUT-charge,  $a<<1$, so that the radial functions
$R_4$ and $R_3$ are only slightly differ from the conditions  (\ref{R34cond}):
$$
R_4=-R_1 + a f_{14}, \qquad R_3=-R_2 + a f_{23}.
$$
Taking this in mind, from eqs.  (\ref{eqsysRgx}) we obtain

\begin{equation}
\begin{split}
\Big( \sqrt{\Delta } \frac{d}{d x}-\frac{i a \sqrt{\Delta } }{\rho^2} - \frac{i \rho
^2)}{\sqrt{\Delta}}\Big) \left(R_2 - a f_{23}\right)   - (i M x + M a+ \Lambda )R_1 =  - a \Lambda f_{14}, \\
 \Big( \sqrt{\Delta } \frac{d}{d x}-\frac{i a
\sqrt{\Delta } }{\rho^2} + \frac{i \rho^2}
{\sqrt{\Delta}}\Big) \left( R_1 - a f_{14}\right) + (i M x + M a + \Lambda ) R_2 =  - a \Lambda f_{23} , \\
  \Big( \sqrt{\Delta } \frac{d}{d x}+\frac{i a
\sqrt{\Delta } }{\rho^2} + \frac{i
\rho^2}{\sqrt{\Delta}}\Big) R_1
 + (i M x - M a -\Lambda ) R_2   = a (i M x - M a) f_{23}, \\
 \Big( \sqrt{\Delta } \frac{d}{d x}+\frac{i a
\sqrt{\Delta } }{\rho^2} - \frac{i
\rho^2}{\sqrt{\Delta}}\Big) R_2 - ( i M x - M a+ \Lambda) R_1
 =  - a (i M x - M a) f_{14}.
\end{split}
\label{eqsysRgx_small-a}
\end{equation}

Removing the right-hand parts in last  two equations at small  $a$,
we arrive at two equations  with  respect to
 $R_1,\,R_2$

\begin{equation}
\begin{split}
\Big( \sqrt{\Delta } \frac{d}{d x}+\frac{i a \sqrt{\Delta } }{\rho^2} + \frac{i \rho^2}{\sqrt{\Delta}}\Big) R_1  + (i M x - M a -\Lambda ) R_2   = 0, \\
\Big( \sqrt{\Delta } \frac{d}{d x}+\frac{i a \sqrt{\Delta } }{\rho^2} - \frac{i
\rho^2}{\sqrt{\Delta}}\Big) R_2  - ( i M x - M a+ \Lambda) R_1  =  0.  \end{split}
\label{eqsysR12_small-a}
\end{equation}
 Equations (\ref{eqsysR12_small-a}) lead to  the 2nd order equations
\begin{equation}
\begin{split}
\Delta R_1 '' + \Big[\frac{2 i a \Delta}{a^2+x^2} +\frac{1}{2}\Delta' + \frac{i M \Delta}{ \Lambda - i M (x+ i a)}
 \Big] R_1' \qquad \qquad \qquad \\ +\Big[ -\Lambda^2+ 2 i x + \frac{
(a^2+x^2)^2}{\Delta}  - \frac{a (a + 2 i x )
\Delta}{(a^2+x^2)^2}   +\frac{i a \Delta'}{2(a^2+x^2)} - \frac{i
 (a^2+x^2) \Delta'}{2 \Delta} \\ - M^2 (x+ i a)^2 -  \frac{M (x^2 + a^2) }{\Lambda - i M (x+ i a)}
 - \frac{a M \Delta}{(\Lambda - i M (x+ i a))(a^2+x^2)} \Big] R_1=0,
\end{split}
\label{eqR1x_small-a}
\end{equation}
\begin{equation}
\begin{split}
\Delta R_2 '' + \Big[\frac{2 i a \Delta}{a^2+x^2} +\frac{1}{2}\Delta' - \frac{i M \Delta}{ \Lambda + i M (x+ i a)}
 \Big] R_2'   \qquad \qquad \qquad \\ +\Big[ -\Lambda^2- 2 i x + \frac{
(a^2+x^2)^2}{\Delta}  - \frac{a (a + 2 i x )
\Delta}{(a^2+x^2)^2}     +\frac{i a \Delta'}{2(a^2+x^2)} + \frac{i
(a^2+x^2) \Delta'}{2 \Delta} \\ - M^2 (x+ i a)^2 -  \frac{M (x^2 + a^2) }{\Lambda + i M (x+ i a)}
 + \frac{a M \Delta}{(\Lambda + i M (x+ i a))(a^2+x^2)}\Big] R_2=0.
\end{split}
\label{eqR2x_small-a}
\end{equation}

For brevity, we will follows only results for  $R_1$.
Applying  in eq. (\ref{eqR1x_small-a}) the substitution  $Z_1 = V_{1a} R_1,$
$$
V_{1a}= i \frac{e^{i x_1} \sqrt{a+i x}
\left(x-x_1\right)^{-1/2-i x_1} (x-x_2)^{-1/2-i x_2}}{\sqrt{a-i x}},
$$
 we obtain the following equation for $Z_1$
\begin{equation}
\begin{split}
Z_1''+\Big(\frac{i (4 x_1-3 i)}{2 (x-x_1)}+\frac{i (4x_2-3 i)}{2
(x-x_2)}- \frac{M}{i \Lambda + M x + i a M } \Big)Z_1'
\\ + \Big( 1-M^2+ \frac{A + B x + C x^2}{(x-x_1)
(x-x_2)(i \Lambda + M x + i a M)}\Big) Z_1=0,
\end{split}
\label{eqZ1_small-a}
\end{equation}
where
$$
A=-i \Lambda \left(8 a^2-2 \Lambda^2+3 i  x_{g}+2\right) - 2 i a M \left(a (4
   a-3)-\Lambda^2+1\right)+ (3 a-1) M  x_{g},
$$ \begin{small} $$  B=- M \left(16 a^2+\left(2 a M^2-4 a-1\right) (2 a-i x_{g})\right)
- i \Lambda \left(-2 M^2
   (x_{g}+2 i a)+4 x_{g}+2 i\right)+2  \Lambda^2 M,
$$ \end{small}
$$
C= 2  M \left( M^2 (2 i a  + x_g) - 2  x_g \right). $$

In the variable $v=(x-x_2)/(x_1-x_2)$, we obtain the simpler description
$$Z_1'' + \Big( \frac{z_2}{v} + \frac{z_1}{v-1} +\frac{-1}{v-c} \Big) Z_1' $$$$+ \Big( 1-M^2+  \frac{(A+B x_2 + C x_2^2)/ (z_1-z_2)+  (B+ 2 C x_2) v + C (z_1-z_2) v^2}{ M v (v-1) (v-c)} \Big)Z_1 =0,
$$
$$c=\frac{i \Lambda +M x_2+i a M }{M (z_2-z_1)}.$$

The last equation  reduces significantly
if
$M=1$ and $C=0$. These  conditions give the following constraint on the parameters
\begin{equation}
\begin{split}
2  M \left( M^2 (2 i a  + x_g) - 2  x_g \right)=0, \; \qquad\mbox{or}\;\qquad
x_g= \frac{2 i a M^2}{2-M^2}, \quad M=1.
\end{split}
\label{xgcond}
\end{equation}

\section{ Extremal NUT black hole}

Let us consider a special case when  $M=1$, then from the condition (\ref{xgcond}) we get $x_g = 2 i a.$
Let us examine imaginary values of  $a$, related to an extremal NUT metric with the imaginary NUT charge.
In this case,  $x_1=x_2=x_g/2$
which  bounds the region of positive values of the  expression under the square  root, $(x_g^2+4 a^2)$.
The Kerr black hole with a single (degenerated) horizon was referred as an extremal Kerr black hole \cite{Aretakis2018}.
By this reason we will use the term "extremal NUT black hole".

Due to equality  $z_1=z_2$ and condition  (\ref{xgcond}),
eq.  (\ref{eqZ1_small-a}) takes the form
\begin{equation}
\begin{split}
Z_1''+\Big(\frac{3-4a}{(x- i a)} - \frac{1}{x - c_{e}} \Big)Z_1'
+ \frac{ b_{e} (-i - c_{e} + x) }{(x - i a)^2 ( x -c_{e} )}Z_1=0,
\end{split}
\label{eqZ1x_small-a-ex}
\end{equation}
where $c_{e} = - i (a+ \Lambda ),$ $b_{e}=(2 a-\Lambda -1)(2 a+\Lambda)$.
In new variable
$$
v_{e}=\frac{x-i a}{c_{e}-i a},
$$
we get
\begin{equation}
\begin{split}
Z_1''+\Big(\frac{3-4a}{v_e} - \frac{1}{v_e - 1} \Big)Z_1' + \frac{ b_{e} (s_{e} + v_e) }{v_e^2 ( v_e -1 )}Z_1=0,
\quad s_e=\frac{1}{a + i c_e}-1. \end{split}
\label{eqZ1v_small-a-ex}
\end{equation}
So, we have an equation of the hypergeometric type, its general solution reads
\begin{equation}
\begin{split}
Z_1=  c_1 \, v_e^{2 a-\alpha _e-1} \, {}_2F_1\Big[ - \alpha _e - \frac{\beta_e}{2}
-\frac{1}{2}, - \alpha _e+ \frac{\beta_e}{2} - \frac{1}{2} ;1-2 \alpha _e;v_e \Big]  \\ + c_2 \, v_e^{2 a+\alpha _e-1}  \,
_2F_1\Big[  \alpha _e-\frac{\beta_e}{2}-\frac{1}{2},  \alpha _e+\frac{\beta_e}{2}
-\frac{1}{2};1+ 2 \alpha _e;v_e\Big],
\end{split}
\label{Z1exM1}
\end{equation}
Taking into account definitions
$$
\alpha _e=\sqrt{(1-2a)^2+ b_e s_e}=\Lambda, \qquad \beta_e=\sqrt{(1-4a)^2-4b_e}=(1+2\Lambda),
$$
 the solution  (\ref{Z1exM1}) simplifies
\begin{equation}
Z_1=   c_1 v^{2 a-\Lambda -1} +c_2 v^{2 a+\Lambda -1} \left(1-\frac{2 \Lambda  v}{2 \Lambda +1}\right).
\end{equation}
Correspondingly, the complete radial function
takes the form
\begin{equation}
\begin{split}
\Delta^{-\frac{1}{4}} \rho_{+}^{-\frac{1}{2}} V_{1a}^{-1} Z_1= C_1  R_{11} +C_2  R_{12}  =C_1 (x-| a| )^{-1-\Lambda} \\
 +C_2
(x-| a| )^{-1+\Lambda}  \left(1+\frac{2 \Lambda ( x- |
   a| )}{\left(1+2 \Lambda\right) \left( i \Lambda + 2  | a|
   \right)}\right),
\end{split}
\label{radNUTextremal}
\end{equation}
where $a=-i |a|$, and according to formula (\ref{quantcondp}) the separation constant equals
 $\Lambda= \pm \sqrt{N(N - 4 i |a| \epsilon)}$.

\begin{figure}[hbt]
\begin{center}
\includegraphics[width=6.5cm,height=4.5cm,angle=0]{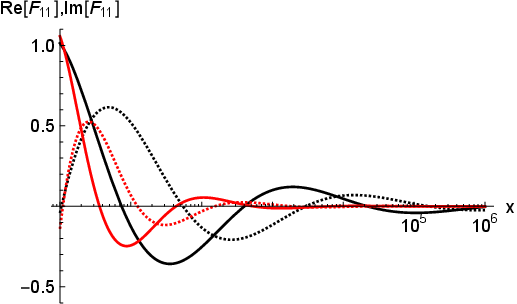}
\includegraphics[width=6.5cm,height=4.5cm,angle=0]{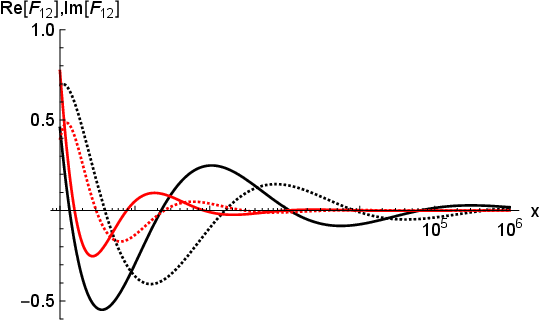}
\end{center}
\caption{The  real (solid lines) and imaginary (dashed lines) parts of the components
$R_{11}$, $R_{12}$ of the wave function (\ref{radNUTextremal})
at the NUT parameter $a=0.05$ (black) and
$a=0.1$ (pink); $N=1, \, \,  \epsilon=10.$ }
\label{figRExtr}
\end{figure}

The behavior of the terms $F_{11} = (x-| a| )^{2} R_{11}$  and $ F_{12}= C_2 (x-| a| ) R_{12}$ is presented in Fig. \ref{figRExtr} (the multipliers attenuate the strong damping and are introduced for better visualization of functions behavior).
As one can see, the increasing of NUT charge $|a|$ leads to the significant  decrease of the $R_{11}$ and $R_{12}$  amplitudes.

At small values of the energy $\epsilon$, we can perform the approximation $\Lambda = \pm \sqrt{N(N - 4 i |a| \epsilon)} \approx \pm N (1 - 2 i |a| \epsilon / N) $.
 Then the real part of
$R_{11}$ can be written as
\begin{equation}
\begin{split}
\mbox{Re}[R_{11}]= \mbox{Re}\Big[(x-| a| )^{-1-\Lambda}\Big]  \approx (x-| a| )^{-1\mp N} \mbox{Re} \Big[e^{\pm 2 i |a| \epsilon \ln{(x-|a|)}}\Big] \quad\\
= \pm (x-| a| )^{-1\mp N} \cos \Big( 2 |a| \epsilon \ln{(x - |a|)}\Big).
\end{split}
\label{exterNUTR11}
\end{equation}

The expansion of the logarithm near any fixed point $x_0$ has the form
\begin{equation}
\begin{split}
 \ln{(x - |a|)} = \ln{\Big((x_0 - |a|)+ (x - x_0)\Big)} \qquad\qquad\\ \approx \frac{1}{x_0-|a|}x +\Big( \ln{(x_0 - |a|)}-\frac{x_0}{x_0-|a|} \Big) \equiv k_{x_0}x +\varphi_{x_0}.
\end{split}
\nonumber
\end{equation}
In other words, in the small vicinity of the point $x_0$ $(x\ll2x_0-|a|)$ the formula (\ref{exterNUTR11}) is rewritten as
$$\mbox{Re}[R_{11}]= \pm (x-| a| )^{-1\mp N} \cos\Big( 2 |a|\epsilon (k_{x_0} \, x +  \varphi_{x_0})\Big).$$
So, one can conclude that the wave number $k_{x_0}$ increases with the rise of
the NUT charge $|a|$; the phase shift $\varphi_{x_0}$ also changes for non-zero NUT charge in comparison with the
Schwarzschild case. These effects are revealed in Fig. \ref{figRExtr}.
These peculiarities may be used to distinguish between Schwarzschild and NUT black holes.

\section{Discussion}

Instead of widely used the Newman-Penrose formalism \cite{NewmanPenrouse1962},
we have applied the usual tetrad Weyl-Fock-Ivanenko method \cite{RedkovNPCS,1928-Tetrode,1929-Fock-Ivanenko,1929-Fock(3),1929-Weyl} to handle with the Dirac equation in NUT space.

It should be noted that radial systems for original NUT  and Taub-NUT spaces are the same. In \cite{5a},  the Dirac equation in the background of Taub-NUT space has been considered and the separation of the variables has been performed by applying the  following substitution:
\begin{equation}
\begin{split}
 X_1=R_{+\frac{1}{2}}(r) T_1(\theta), \,  X_2=R_{-\frac{1}{2}}(r)
T_2(\theta), \, Y_1=R_{-\frac{1}{2}}(r) T_1(\theta), \,
Y_2=R_{+\frac{1}{2}}(r) T_2(\theta),
\end{split}
\label{subwrong}
\end{equation}
 which previously successfully used for separating the variables  in the Dirac equation in the background of Kerr and Kerr-Newman spacetimes \cite{Dariescu2021,Chanrdasecar1976}.
However, being applied to Dirac problem in NUT space, this substitution gives the inconsistent system of four equations. Indeed, applying the substitution (\ref{subwrong}) to the equations (\ref{DiracXY}), one get
\begin{equation}
\begin{split}
   \Big( \sqrt{\Delta }
\frac{\partial}{\partial r}-\frac{i a \sqrt{\Delta } }{\rho ^2} -
\frac{i \epsilon \rho^2}{\sqrt{\Delta}} \Big) R_{-\frac{1}{2}}   +
i M (r-ia)R_{+\frac{1}{2}} = \Lambda R_{+\frac{1}{2}} ,
\\
   \Big( \sqrt{\Delta }
\frac{\partial}{\partial r}+\frac{i a \sqrt{\Delta } }{\rho ^2} -
\frac{i \epsilon \rho^2}{\sqrt{\Delta}}\Big) R_{-\frac{1}{2}}  + i
M (r+ia)R_{+\frac{1}{2}} = \Lambda R_{+\frac{1}{2}},
\\
    \Big( \sqrt{\Delta }
\frac{\partial}{\partial r}+\frac{i a \sqrt{\Delta } }{\rho ^2} +
\frac{i \epsilon \rho^2}{\sqrt{\Delta}}\Big) R_{+\frac{1}{2}}   -
i M (r+ia)R_{-\frac{1}{2}} = \Lambda R_{-\frac{1}{2}},
\\
   \Big( \sqrt{\Delta }
\frac{\partial}{\partial r}-\frac{i a \sqrt{\Delta } }{\rho ^2} +
\frac{i \epsilon \rho^2}{\sqrt{\Delta}}\Big) R_{+\frac{1}{2}}   -
i M (r-ia)R_{-\frac{1}{2}} = \Lambda R_{-\frac{1}{2}}. \end{split}
\label{Dirwrong}
\end{equation}

Subtracting the first equation from the second, and also the fourth equation from the third,
one obtain algebraic relations
$$\frac{2ia \sqrt{\Delta}}{\rho^2}R_{-\frac{1}{2}}- 2 M a R_{+\frac{1}{2}}=0, \quad
\frac{2ia \sqrt{\Delta}}{\rho^2}R_{+\frac{1}{2}}+ 2 M a R_{-\frac{1}{2}}=0,$$
the last system leads to
$$\left(\frac{\Delta}{\rho^4} -M^2 \right) R_{+\frac{1}{2}}=0, \quad
\left(\frac{\Delta}{\rho^4} - M^2 \right) R_{-\frac{1}{2}}=0.$$
Because $\left(\frac{\Delta^2}{\rho^4} -M^2 \right)\neq0$ for
any radial coordinate $r$, we conclude that the unique solution for the system  (\ref{Dirwrong}) is trivial,
$R_{+\frac{1}{2}}=R_{-\frac{1}{2}}\equiv 0$.

From such inconsistent system of four equations, the authors of \cite{5a} has  derived  the new system  by linear combination, but
have preserved only two equations instead of four. Obviously, the solution  of such two-equation system is not solution of the initial system of four equations.  By this reason we consider the result of \cite{5a} relating to the radial system of Dirac equation as doubtful.

In our study we have proved that for the quantum-mechanical problem  of a
massive particle in the NUT spacetime it is impossible to
rearrange the four radial equations by applying the linear
constraints for radial functions, to reduce the problem to two
first order differential equations, for details see Supplement B. Because of that, we analyze the system of four linked differential equations
(\ref{Direqseparation a})  and solve it for massless particle.
Recently \cite{KrylovaNPCS2024}, solutions of the Dirac equation for a massless particle have been described
within the Frobenius approach. In the present paper, we have constructed the solution in terms of the confluent Heun functions.
In contrast to the Frobenius-type solutions, the Heun confluent functions allow to get the scattering
resonance frequencies \cite{Vieira2016}. We have shown that the presence of NUT charge leads
to the decreasing resonance energies compared with the Schwarzschild black hole.

For the NUT metric,  the Riemann curvature tensor turns to zero at infinity, but
the presence of the Misner string makes the
NUT spacetime asymptotically non-flat and anisotrophic \cite{3}.
Though the NUT metric possesses the string singularity and, correspondingly,  there exist
closed timelike curves, in \cite{Clement2016}  it was shown that
the geodesics of the freely falling observers are not closed timelike curves.
So, NUT metric avoids causality violation, and it can be considered as physically meaningful.
By this reason, the primordial black hole with NUT charge seems to be an intriguing object
in early Universe models \cite{Chakraborty2022}. As shown in \cite{Chakraborty2022},
the low-energy (related to the ordinary mass less than 5$\times10^{11}$ kg) primordial black hole
 with NUT charge 
have the smaller Hawking temperature and may not be decayed due to the Hawking radiation by now.

In \cite{Chakraborty2022,Liu2022}, the Hawking radiation was defined in the conventional way, in contrast to this, in the present paper
we have found this temperature from the structure of the Dirac equation solutions, and have demonstrated the decreasing of the probability of
particle-antiparticle  production on the event horizon. This supports the result of \cite{Chakraborty2022}, that
the primordial black holes with large NUT charge,  could be preserved in the present Universe.
As known, the Hawking radiation even from the Schwarzschild black hole was not observed till now. But if such experiments will exist in  future, they may serve an
experimental test for distinguishing black holes of different types. Let compare the Hawking radiation for the NUT and Kerr black holes, the last is much more examined.
As it was shown in \cite{Dariescu2021},
the Hawking temperature for the Kerr black hole is determined by the formula (with the notations used in our paper $r_g=2M_{BH}$, $M_{BH}$ is a mass of black hole):
$$T_{\mbox{Kerr}}=\frac{\sqrt{r_g^2-4J^2}}{2\pi r_g(r_g+\sqrt{r_g^2-4J^2})},$$
here $J$ is the angular momentum per unit mass.
For NUT black hole, we correspondingly have
$$T_{\mbox{NUT}}=\frac{1}{2\pi (r_g+\sqrt{r_g^2+4a^2})}.$$ As one can see in Fig. \ref{figT} the rotation of black hole as well as NUT charge result in decrease of the Hawking temperature,
but the characters of these effects  are different.

\begin{figure}[hbt]
\begin{center}
\includegraphics[width=6.5cm,height=4.5cm,angle=0]{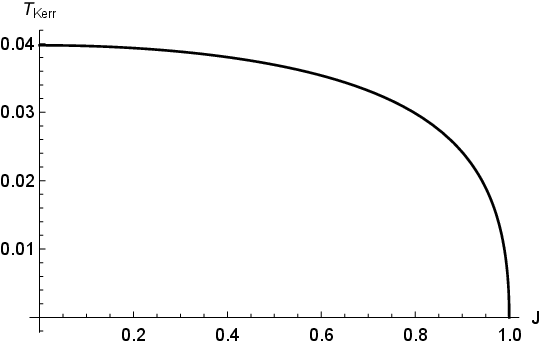}  \hspace{1.0cm}
\includegraphics[width=6.5cm,height=4.5cm,angle=0]{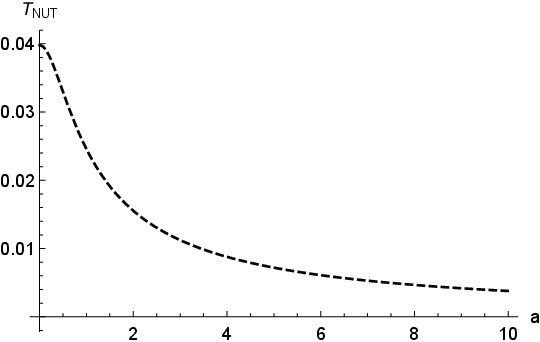}
\end{center}
\caption{The  Hawking temperature for Kerr (left)  and NUT (right)  black holes in dependence on the angular momentum $J$ and NUT-parameter $a$, respectively. }
\label{figT}
\end{figure}

We have noted that the imaginary NUT parameter $a$ leads to complex-valued metric (\ref{nutmetr}).
As known in quantum gravity models, the complex-valued metrics may be able to regularise the big bang
by using a non-singular geometry.  However, not all complex metrics may be considered as
physically interpretable, the relevant criteria of applicability are described in \cite{Jonas2022,Briscese2022}.

Recall that in quantum mechanics the use of the complex wave functions is admissible when the observables
quantities are real. Interestingly, that the effective potential for massless particle
is complex-valued for real NUT parameter. However, at imaginary NUT charge
the effective potential is real-valued (Section 5.2.1).

In \cite{Gibbons2017}, it was shown that re-scaling of the vacuum Weyl metrics 
and transforming it to a complex parameter yields axially symmetric vacuum
solutions 
with wormhole topology; their sources can be viewed as thin rings of the negative
tension encircling the throats. Such ring wormholes do not show infinite red or blue shifts as well as pathologies
like closed timelike curves. The supercritically charged black holes with NUT parameter belong to
these traversable wormholes \cite{Clement2016}.
For pure (uncharged) black hole with a real NUT parameter, the wormhole solutions do not exist.
However, when using imaginary NUT parameter values one can obtain a wormhole solution.

According  \cite{Hawking1995}, the entropy of an extremal black hole should vanish, though
its event horizon can have nonzero area. For instance, the Bekenstein-Hawking entropy formula $S=A/4$ cannot
be applied to the extremal case for Kerr and  Reissner-Nordstr\"{o}m solutions.
Because of that,  transition to extremality for these spacetime models is not continuous  \cite{Carroll2009}.
For extremal NUT black hole (Section 7), we face the opposite situation.
The entropy of the NUT black hole  equals   $S=\pi (a^2+r_2^2)$ \cite{Liu2022}, so it vanishes $S=0$ if  $r_2=i a$.
This fact makes this complex-valued metric to be interesting  for theoretical investigation.

\section{Conclusion}

Therefore, one can conclude that the Dirac equation for the particle in the background of the Newman-Unti-Tamburino spacetime has been derived by applying the tetrad formalism.
The separation procedure has been performed and the general wave function substitution (\ref{PsiL}) has been found. The obtained two different modes point out on the existence of additional operator of the helicity type. The finding of the explicit form of this operator is a task of the future investigations.

The system of two differential
equations for angular functions and the system of four
differential equations for radial functions have been derived. The angular equations have been solved in terms of hypergeometric functions and  the NUT-charge dependent  quantization rule for the angular separation constant have been found.
The radial system has been solved for special cases of the massless particle and for massive particle with a fixed energy near extremal NUT black hole.  For the massless fermion, we have been constructed the solution of the radial system of Dirac equation in terms of the confluent Heun functions.  We have demonstrated that, in opposite to the case of Kerr black hole, the values of resonant energies do not possess the real part and are imaginary ones, similarly to the Schwarzschild case. It seems to indicate the absence of the bound states. Nevertheless, the non-vanishing NUT charge influences the wave characteristics, for example, results in decrease of resonant energies absolute values.

We have demonstrated that the potentials become complex-valued functions at real NUT charge, but, interestingly,
they become real-valued  at imaginary NUT-charge. The last fact has encouraged us to consider the extremal NUT black hole with a single horizon, when  the
Bekenstein-Hawking entropy vanishes identically, and reveal the non-zero NUT charge effects in wave characteristics. We have shown that the wave number increases with the rise of the NUT charge.

As a result of analysis of the
radial equations for massive particle, it has been demonstrated that similarly to Kerr black hole, the probability of particle-antiparticle production on the outer
event horizon decreases for NUT black hole comparing to Schwarzschild case,  but the characters of the effects of angular moment (for Kerr black hole) and NUT charge (for NUT black hole) are different.

The results obtained in the paper may distinguish NUT black holes from another type of black holes in the future when the experimental setup will allow such type of observations. But today, there arise the necessity of the following elucidation and explanation of the NUT charge from the physical point of view and  finding an appropriate constraints on its values.


\begin{thebibliography}{41}
\expandafter\ifx\csname
natexlab\endcsname\relax\def\natexlab#1{#1}\fi
\expandafter\ifx\csname bibnamefont\endcsname\relax
  \def\bibnamefont#1{#1}\fi
\expandafter\ifx\csname bibfnamefont\endcsname\relax
  \def\bibfnamefont#1{#1}\fi
\expandafter\ifx\csname citenamefont\endcsname\relax
  \def\citenamefont#1{#1}\fi
\expandafter\ifx\csname url\endcsname\relax
  \def\url#1{\texttt{#1}}\fi
\expandafter\ifx\csname
urlprefix\endcsname\relax\def\urlprefix{URL }\fi
\providecommand{\bibinfo}[2]{#2}
\providecommand{\eprint}[2][]{\url{#2}}

\bibitem[{\citenamefont{Manko and Ruiz}(2005)}]{Manko2005}
\bibinfo{author}{\bibfnamefont{V.~S.} \bibnamefont{Manko}} \bibnamefont{and}
  \bibinfo{author}{\bibfnamefont{E.}~\bibnamefont{Ruiz}},
  \bibinfo{journal}{Class. Quantum Grav.} \textbf{\bibinfo{volume}{22}},
  \bibinfo{pages}{3555} (\bibinfo{year}{2005}).

\bibitem[{\citenamefont{Newman et~al.}(1963)\citenamefont{Newman, Tamburino,
  and Unti.}}]{Newman}
\bibinfo{author}{\bibfnamefont{E.}~\bibnamefont{Newman}},
  \bibinfo{author}{\bibfnamefont{L.}~\bibnamefont{Tamburino}},
  \bibnamefont{and} \bibinfo{author}{\bibfnamefont{T.}~\bibnamefont{Unti.}},
  \bibinfo{journal}{J.Math.Phys.} \textbf{\bibinfo{volume}{4}},
  \bibinfo{pages}{915} (\bibinfo{year}{1963}).

\bibitem[{\citenamefont{Chakraborty and Mukhopadhyay}(2023)}]{Chakraborty2023}
\bibinfo{author}{\bibfnamefont{C.}~\bibnamefont{Chakraborty}} \bibnamefont{and}
  \bibinfo{author}{\bibfnamefont{B.}~\bibnamefont{Mukhopadhyay}},
  \bibinfo{journal}{Eur. Phys. J. C} \textbf{\bibinfo{volume}{83}},
  \bibinfo{pages}{937} (\bibinfo{year}{2023}).

\bibitem[{\citenamefont{Ghasemi-Nodehi}(2024)}]{Ghasemi-Nodehi2024}
\bibinfo{author}{\bibfnamefont{M.}~\bibnamefont{Ghasemi-Nodehi}},
  \bibinfo{journal}{Universe} \textbf{\bibinfo{volume}{10}},
  \bibinfo{pages}{378} (\bibinfo{year}{2024}).

\bibitem[{\citenamefont{Ghasemi-Nodehi
  et~al.}(2021)\citenamefont{Ghasemi-Nodehi, Chakraborty, Yu, and Lu}}]{3}
\bibinfo{author}{\bibfnamefont{M.}~\bibnamefont{Ghasemi-Nodehi}},
  \bibinfo{author}{\bibfnamefont{C.}~\bibnamefont{Chakraborty}},
  \bibinfo{author}{\bibfnamefont{Q.}~\bibnamefont{Yu}}, \bibnamefont{and}
  \bibinfo{author}{\bibfnamefont{Y.}~\bibnamefont{Lu}}, \bibinfo{journal}{Eur.
  Phys. J. C} \textbf{\bibinfo{volume}{81}}, \bibinfo{pages}{939}
  (\bibinfo{year}{2021}).

\bibitem[{\citenamefont{Liu et~al.}(2022)\citenamefont{Liu, L$\ddot{u}$, and
  Ma}}]{Liu2022}
\bibinfo{author}{\bibfnamefont{H.}~\bibnamefont{Liu}},
  \bibinfo{author}{\bibfnamefont{H.}~\bibnamefont{L$\ddot{u}$}},
  \bibnamefont{and} \bibinfo{author}{\bibfnamefont{L.}~\bibnamefont{Ma}},
  \bibinfo{journal}{J. High Energ. Phys.} \textbf{\bibinfo{volume}{2022}},
  \bibinfo{pages}{174} (\bibinfo{year}{2022}).

\bibitem[{\citenamefont{Awad and Elkhateeb}(2023)}]{4}
\bibinfo{author}{\bibfnamefont{A.}~\bibnamefont{Awad}} \bibnamefont{and}
  \bibinfo{author}{\bibfnamefont{E.}~\bibnamefont{Elkhateeb}},
  \bibinfo{journal}{Phys. Rev. D} \textbf{\bibinfo{volume}{108}},
  \bibinfo{pages}{064022} (\bibinfo{year}{2023}).

\bibitem[{\citenamefont{Atiyah and Hitchin}(1985)}]{Atiyah}
\bibinfo{author}{\bibfnamefont{M.~F.} \bibnamefont{Atiyah}} \bibnamefont{and}
  \bibinfo{author}{\bibfnamefont{N.}~\bibnamefont{Hitchin}},
  \bibinfo{journal}{Phys. Lett. A} \textbf{\bibinfo{volume}{107}},
  \bibinfo{pages}{21} (\bibinfo{year}{1985}).

\bibitem[{\citenamefont{Cot\u{a}escu and Visinescu}(2001)}]{4a}
\bibinfo{author}{\bibfnamefont{I.~I.} \bibnamefont{Cot\u{a}escu}}
  \bibnamefont{and}
  \bibinfo{author}{\bibfnamefont{M.}~\bibnamefont{Visinescu}},
  \bibinfo{journal}{Int. J. Mod. Phys.} \textbf{\bibinfo{volume}{16}},
  \bibinfo{pages}{1743} (\bibinfo{year}{2001}).

\bibitem[{\citenamefont{Godazgar et~al.}(2019)\citenamefont{Godazgar, Godazgar,
  and Pope}}]{Godazgar}
\bibinfo{author}{\bibfnamefont{H.}~\bibnamefont{Godazgar}},
  \bibinfo{author}{\bibfnamefont{M.}~\bibnamefont{Godazgar}}, \bibnamefont{and}
  \bibinfo{author}{\bibfnamefont{C.}~\bibnamefont{Pope}},
  \bibinfo{journal}{Phys. Lett. B} \textbf{\bibinfo{volume}{798}},
  \bibinfo{pages}{134938} (\bibinfo{year}{2019}).

\bibitem[{\citenamefont{Zimmerman and Shahir}(1989)}]{Zimmerman1989}
\bibinfo{author}{\bibfnamefont{R.~L.} \bibnamefont{Zimmerman}}
  \bibnamefont{and} \bibinfo{author}{\bibfnamefont{B.}~\bibnamefont{Shahir}},
  \bibinfo{journal}{Gen. Relativ. Gravit.} \textbf{\bibinfo{volume}{21}},
  \bibinfo{pages}{821} (\bibinfo{year}{1989}).

\bibitem[{\citenamefont{Nouri-Zonoz and Lynden-Bell}(1997)}]{Nouri-Zonoz1997}
\bibinfo{author}{\bibfnamefont{M.}~\bibnamefont{Nouri-Zonoz}} \bibnamefont{and}
  \bibinfo{author}{\bibfnamefont{D.}~\bibnamefont{Lynden-Bell}},
  \bibinfo{journal}{Mon. Not. R. Astron. Soc.} \textbf{\bibinfo{volume}{292}},
  \bibinfo{pages}{714} (\bibinfo{year}{1997}).

\bibitem[{\citenamefont{Vandeev and Semenova}(2022)}]{Vandeev2022}
\bibinfo{author}{\bibfnamefont{V.~P.} \bibnamefont{Vandeev}} \bibnamefont{and}
  \bibinfo{author}{\bibfnamefont{A.~N.} \bibnamefont{Semenova}},
  \bibinfo{journal}{Int. J. Modern Phys. D} \textbf{\bibinfo{volume}{31}},
  \bibinfo{pages}{2250108} (\bibinfo{year}{2022}).

\bibitem[{\citenamefont{Nouri-Zonoz}(2004)}]{Nouri-Zonoz2004}
\bibinfo{author}{\bibfnamefont{M.}~\bibnamefont{Nouri-Zonoz}},
  \bibinfo{journal}{Class. Quantum Grav.} \textbf{\bibinfo{volume}{21}},
  \bibinfo{pages}{471} (\bibinfo{year}{2004}).

\bibitem[{\citenamefont{Al-Badawi et~al.}(2022)\citenamefont{Al-Badawi, Kanzi,
  and Sakalli}}]{5a}
\bibinfo{author}{\bibfnamefont{A.}~\bibnamefont{Al-Badawi}},
  \bibinfo{author}{\bibfnamefont{S.}~\bibnamefont{Kanzi}}, \bibnamefont{and}
  \bibinfo{author}{\bibfnamefont{I.}~\bibnamefont{Sakalli}},
  \bibinfo{journal}{Eur. Phys. J. Plus} \textbf{\bibinfo{volume}{137}},
  \bibinfo{pages}{94} (\bibinfo{year}{2022}).

\bibitem[{\citenamefont{Simulik}(2025)}]{Simulik2025}
\bibinfo{author}{\bibfnamefont{V.}~\bibnamefont{Simulik}}, \bibinfo{journal}{J.
  Phys. A: Math. Theor.} \textbf{\bibinfo{volume}{58}}, \bibinfo{pages}{053001}
  (\bibinfo{year}{2025}).

\bibitem[{\citenamefont{Simulik}(2020)}]{Simulik2020}
\bibinfo{author}{\bibfnamefont{V.}~\bibnamefont{Simulik}},
  \emph{\bibinfo{title}{Relativistic Quantum Mechanics and Field Theory of
  Arbitrary Spin}} (\bibinfo{publisher}{Nova Science, New York.},
  \bibinfo{year}{2020}).

\bibitem[{\citenamefont{Tetrode}(1928)}]{1928-Tetrode}
\bibinfo{author}{\bibfnamefont{H.}~\bibnamefont{Tetrode}},
  \bibinfo{journal}{Zeit. Phys.} \textbf{\bibinfo{volume}{50}},
  \bibinfo{pages}{336} (\bibinfo{year}{1928}).

\bibitem[{\citenamefont{Weyl}(1929)}]{1929-Weyl}
\bibinfo{author}{\bibfnamefont{H.}~\bibnamefont{Weyl}}, \bibinfo{journal}{Proc.
  Nat. Acad. Sci. Amer.} \textbf{\bibinfo{volume}{15}}, \bibinfo{pages}{323}
  (\bibinfo{year}{1929}).

\bibitem[{\citenamefont{Fock and Ivanenko}(1929)}]{1929-Fock-Ivanenko}
\bibinfo{author}{\bibfnamefont{V.}~\bibnamefont{Fock}} \bibnamefont{and}
  \bibinfo{author}{\bibfnamefont{D.}~\bibnamefont{Ivanenko}},
  \bibinfo{journal}{Zeit. Phys.} \textbf{\bibinfo{volume}{54}},
  \bibinfo{pages}{798} (\bibinfo{year}{1929}).

\bibitem[{\citenamefont{Fock}(1929)}]{1929-Fock(3)}
\bibinfo{author}{\bibfnamefont{V.}~\bibnamefont{Fock}}, \bibinfo{journal}{Zeit.
  Phys.} \textbf{\bibinfo{volume}{57}}, \bibinfo{pages}{261}
  (\bibinfo{year}{1929}).

\bibitem[{\citenamefont{Newman and Penrose}(1962)}]{NewmanPenrouse1962}
\bibinfo{author}{\bibfnamefont{E.}~\bibnamefont{Newman}} \bibnamefont{and}
  \bibinfo{author}{\bibfnamefont{R.}~\bibnamefont{Penrose}},
  \bibinfo{journal}{J. Math. Phys.} \textbf{\bibinfo{volume}{3}},
  \bibinfo{pages}{566} (\bibinfo{year}{1962}).

\bibitem[{\citenamefont{Landau and Lifshitz}(1973)}]{Landau}
\bibinfo{author}{\bibfnamefont{L.}~\bibnamefont{Landau}} \bibnamefont{and}
  \bibinfo{author}{\bibfnamefont{E.}~\bibnamefont{Lifshitz}},
  \emph{\bibinfo{title}{The classical theory of fields}}
  (\bibinfo{publisher}{Nauka, Moscow}, \bibinfo{year}{1973}).

\bibitem[{\citenamefont{Dariescu et~al.}(2021)\citenamefont{Dariescu, Dariescu,
  and Stelea}}]{Dariescu2021}
\bibinfo{author}{\bibfnamefont{C.}~\bibnamefont{Dariescu}},
  \bibinfo{author}{\bibfnamefont{M.-A.} \bibnamefont{Dariescu}},
  \bibnamefont{and} \bibinfo{author}{\bibfnamefont{C.}~\bibnamefont{Stelea}},
  \bibinfo{journal}{Advances in High Energy Phys.}
  \textbf{\bibinfo{volume}{2021}}, \bibinfo{pages}{5512735}
  (\bibinfo{year}{2021}).

\bibitem[{\citenamefont{Sannan}(1988)}]{Sannan1988}
\bibinfo{author}{\bibfnamefont{S.}~\bibnamefont{Sannan}},
  \bibinfo{journal}{Gen. Relativ. Gravit.} \textbf{\bibinfo{volume}{20}},
  \bibinfo{pages}{239} (\bibinfo{year}{1988}).

\bibitem[{\citenamefont{Vieira et~al.}(2015)\citenamefont{Vieira, Bezerra, and
  Silva}}]{Vieira2015}
\bibinfo{author}{\bibfnamefont{H.}~\bibnamefont{Vieira}},
  \bibinfo{author}{\bibfnamefont{V.}~\bibnamefont{Bezerra}}, \bibnamefont{and}
  \bibinfo{author}{\bibfnamefont{G.}~\bibnamefont{Silva}},
  \bibinfo{journal}{Annals of Physics} \textbf{\bibinfo{volume}{362}},
  \bibinfo{pages}{576} (\bibinfo{year}{2015}).

\bibitem[{\citenamefont{Fiziev}(2010)}]{Fiziev2010}
\bibinfo{author}{\bibfnamefont{P.}~\bibnamefont{Fiziev}}, \bibinfo{journal}{J.
  Phys. A: Math. Theor.} \textbf{\bibinfo{volume}{43}}, \bibinfo{pages}{035203}
  (\bibinfo{year}{2010}).

\bibitem[{\citenamefont{Slavyanov and Lay}(2000)}]{Slavyanov2000}
\bibinfo{author}{\bibfnamefont{S.}~\bibnamefont{Slavyanov}} \bibnamefont{and}
  \bibinfo{author}{\bibfnamefont{W.}~\bibnamefont{Lay}},
  \emph{\bibinfo{title}{Special functions. A unified theory based on
  singularities}} (\bibinfo{publisher}{Oxford University Press,Inc., New
  York.}, \bibinfo{year}{2000}).

\bibitem[{\citenamefont{Vieira and Bezerra}(2016)}]{Vieira2016}
\bibinfo{author}{\bibfnamefont{H.}~\bibnamefont{Vieira}} \bibnamefont{and}
  \bibinfo{author}{\bibfnamefont{V.}~\bibnamefont{Bezerra}},
  \bibinfo{journal}{Annals of Physics} \textbf{\bibinfo{volume}{373}},
  \bibinfo{pages}{28} (\bibinfo{year}{2016}).

\bibitem[{\citenamefont{Chichurin et~al.}(2024)\citenamefont{Chichurin,
  Ovsiyuk, and Redkov}}]{5}
\bibinfo{author}{\bibfnamefont{A.}~\bibnamefont{Chichurin}},
  \bibinfo{author}{\bibfnamefont{E.}~\bibnamefont{Ovsiyuk}}, \bibnamefont{and}
  \bibinfo{author}{\bibfnamefont{V.}~\bibnamefont{Redkov}},
  \bibinfo{journal}{Rom. Rep. Phys.} \textbf{\bibinfo{volume}{76}},
  \bibinfo{pages}{110} (\bibinfo{year}{2024}).

\bibitem[{\citenamefont{Aretakis}(2018)}]{Aretakis2018}
\bibinfo{author}{\bibfnamefont{S.}~\bibnamefont{Aretakis}}
  (\bibinfo{publisher}{Springer, Cham.}, \bibinfo{year}{2018}),
  vol.~\bibinfo{volume}{33} of \emph{\bibinfo{series}{SpringerBriefs in
  Mathematical Physic}}, chap. \bibinfo{chapter}{Extremal {K}err Black Holes},
  pp. \bibinfo{pages}{71--81}.

\bibitem[{\citenamefont{Ovsiyuk et~al.}(2017)\citenamefont{Ovsiyuk, Veko,
  Rusak, Chichurin, and Red'kov}}]{RedkovNPCS}
\bibinfo{author}{\bibfnamefont{E.}~\bibnamefont{Ovsiyuk}},
  \bibinfo{author}{\bibfnamefont{O.}~\bibnamefont{Veko}},
  \bibinfo{author}{\bibfnamefont{Y.}~\bibnamefont{Rusak}},
  \bibinfo{author}{\bibfnamefont{A.}~\bibnamefont{Chichurin}},
  \bibnamefont{and} \bibinfo{author}{\bibfnamefont{V.}~\bibnamefont{Red'kov}},
  \bibinfo{journal}{Int. J. Nonlinear. Phenom. Complex. Sys.}
  \textbf{\bibinfo{volume}{20}}, \bibinfo{pages}{56} (\bibinfo{year}{2017}).

\bibitem[{\citenamefont{Chandrasekhar}(1976)}]{Chanrdasecar1976}
\bibinfo{author}{\bibfnamefont{S.}~\bibnamefont{Chandrasekhar}},
  \bibinfo{journal}{Proc. R. Soc. Lond. A} \textbf{\bibinfo{volume}{349}},
  \bibinfo{pages}{571} (\bibinfo{year}{1976}).

\bibitem[{\citenamefont{Krylova and Red'kov}(2024)}]{KrylovaNPCS2024}
\bibinfo{author}{\bibfnamefont{N.}~\bibnamefont{Krylova}} \bibnamefont{and}
  \bibinfo{author}{\bibfnamefont{V.}~\bibnamefont{Red'kov}}, in
  \emph{\bibinfo{booktitle}{Nonlinear Dynamics and Applications: Proc. the 31
  Annual Seminar NPCS'2024}}, edited by
  \bibinfo{editor}{\bibfnamefont{V.}~\bibnamefont{Schaporov}}
  (\bibinfo{organization}{Joint Institute for Power and Nuclear Research –
  Sosny}, \bibinfo{address}{Minsk}, \bibinfo{year}{2024}).

\bibitem[{\citenamefont{Cl\'ement et~al.}(2016)\citenamefont{Cl\'ement,
  Gal'tsov, and Guenouche}}]{Clement2016}
\bibinfo{author}{\bibfnamefont{G.}~\bibnamefont{Cl\'ement}},
  \bibinfo{author}{\bibfnamefont{D.}~\bibnamefont{Gal'tsov}}, \bibnamefont{and}
  \bibinfo{author}{\bibfnamefont{M.}~\bibnamefont{Guenouche}},
  \bibinfo{journal}{Phys. Rev. D} \textbf{\bibinfo{volume}{93}},
  \bibinfo{pages}{024048} (\bibinfo{year}{2016}).

\bibitem[{\citenamefont{Chakraborty and Bhattacharyya}(2022)}]{Chakraborty2022}
\bibinfo{author}{\bibfnamefont{C.}~\bibnamefont{Chakraborty}} \bibnamefont{and}
  \bibinfo{author}{\bibfnamefont{S.}~\bibnamefont{Bhattacharyya}},
  \bibinfo{journal}{Phys. Rev. D} \textbf{\bibinfo{volume}{106}},
  \bibinfo{pages}{103028} (\bibinfo{year}{2022}).

\bibitem[{\citenamefont{Jonas et~al.}(2022)\citenamefont{Jonas, Lehners, and
  Quintin}}]{Jonas2022}
\bibinfo{author}{\bibfnamefont{C.}~\bibnamefont{Jonas}},
  \bibinfo{author}{\bibfnamefont{J.}~\bibnamefont{Lehners}}, \bibnamefont{and}
  \bibinfo{author}{\bibfnamefont{J.}~\bibnamefont{Quintin}},
  \bibinfo{journal}{J. High Energ. Phys.} \textbf{\bibinfo{volume}{2022}},
  \bibinfo{pages}{284} (\bibinfo{year}{2022}).

\bibitem[{\citenamefont{Briscese}(2022)}]{Briscese2022}
\bibinfo{author}{\bibfnamefont{F.}~\bibnamefont{Briscese}},
  \bibinfo{journal}{Phys. Rev. D} \textbf{\bibinfo{volume}{105}},
  \bibinfo{pages}{126028} (\bibinfo{year}{2022}).

\bibitem[{\citenamefont{Gibbons and Volkov}(2017)}]{Gibbons2017}
\bibinfo{author}{\bibfnamefont{G.}~\bibnamefont{Gibbons}} \bibnamefont{and}
  \bibinfo{author}{\bibfnamefont{M.}~\bibnamefont{Volkov}},
  \bibinfo{journal}{JCAP} \textbf{\bibinfo{volume}{1705}}, \bibinfo{pages}{039}
  (\bibinfo{year}{2017}).

\bibitem[{\citenamefont{Hawking et~al.}(1995)\citenamefont{Hawking, Horowitz,
  and Ross}}]{Hawking1995}
\bibinfo{author}{\bibfnamefont{S.~W.} \bibnamefont{Hawking}},
  \bibinfo{author}{\bibfnamefont{G.~T.} \bibnamefont{Horowitz}},
  \bibnamefont{and} \bibinfo{author}{\bibfnamefont{S.~F.} \bibnamefont{Ross}},
  \bibinfo{journal}{Phys. Rev. D} \textbf{\bibinfo{volume}{51}},
  \bibinfo{pages}{4302} (\bibinfo{year}{1995}).

\bibitem[{\citenamefont{Carroll et~al.}(2009)\citenamefont{Carroll, Johnson,
  and Randall}}]{Carroll2009}
\bibinfo{author}{\bibfnamefont{S.}~\bibnamefont{Carroll}},
  \bibinfo{author}{\bibfnamefont{M.~C.} \bibnamefont{Johnson}},
  \bibnamefont{and} \bibinfo{author}{\bibfnamefont{L.}~\bibnamefont{Randall}},
  \bibinfo{journal}{JHEP} \textbf{\bibinfo{volume}{11}}, \bibinfo{pages}{109}
  (\bibinfo{year}{2009}).

\end{thebibliography}

\end{document}